\documentclass{aa}  

\usepackage{graphicx}
\usepackage{txfonts}
\usepackage{graphicx}
\usepackage{lscape}
\usepackage{longtable}
\usepackage{natbib}
\usepackage{color}
\usepackage{array} 
\usepackage{array} 
\usepackage{tikz,array}
\usetikzlibrary{calc}
\usepackage{txfonts}
\usepackage{color}
\usepackage{multirow}
\usepackage{here}

\usepackage{txfonts}
\usepackage{amsmath,amstext}
\usepackage{graphicx}
\usepackage{epstopdf}
\usepackage{float}
\usepackage{array}
\usepackage{mathtools}
\usepackage{booktabs}
\usepackage{subfigure}
\usepackage{url}
\usepackage{helvet}
\usepackage{tabularx}
\usepackage{multirow}
\usepackage{natbib}
\usepackage[flushleft]{threeparttable}
\usepackage{lscape}
\usepackage{pdflscape}
\usepackage{longtable}
\usepackage{wasysym}
\usepackage{float}
\usepackage{relsize}
\usepackage{color}
\usepackage{breqn}
\usepackage{bm}

\usepackage[colorlinks, citecolor=blue, linkcolor=blue]{hyperref}
\hypersetup{colorlinks,breaklinks, linkcolor=blue,urlcolor=magenta, anchorcolor=blue,citecolor=blue}

\newcommand{\be}{\begin{equation}}
\newcommand{\ee}{\end{equation}}

\begin{document}

   \title{Follow-up of non-transiting planets detected by Kepler\thanks{Based on observations collected at Centro Astron\'omico Hispano en Andaluc\'ia (CAHA) at Calar Alto, operated jointly by Instituto de Astrof\'isica de Andaluc\'ia (CSIC) and Junta de Andaluc\'ia and  observations made with the Mercator Telescope, operated on the island of La Palma by the Flemish Community, at the Spanish Observatorio del Roque de los Muchachos of the Instituto de Astrofísica de Canarias.}}
   \subtitle{Confirmation of three hot-Jupiters and validation of three other planets}

   \author{
J. Lillo-Box\inst{\ref{cab}},  
S. Millholland\inst{\ref{Princeton}, \ref{Yale}} \& 
G. Laughlin\inst{\ref{Yale}}
}

\institute{
Centro de Astrobiolog\'ia (CAB, CSIC-INTA), Depto. de Astrof\'isica, ESAC campus 28692 Villanueva de la Ca\~nada (Madrid), Spain\label{cab} \email{Jorge.Lillo@cab.inta-csic.es  }
\and Department of Astrophysical Sciences, Princeton University, Princeton, NJ 08544, USA \label{Princeton}
\and Department of Astronomy, Yale University, New Haven, CT 06511, USA 
\label{Yale}
}

\titlerunning{Follow-up of non-transiting planets detected by Kepler}
\authorrunning{Lillo-Box et al.}

   \date{In preparation}

 
  \abstract
   {The direct detection of new extrasolar planets from high-precision photometry data is commonly based on the observation of the transit signal of the planet as it passes in front of its star. Close-in planets, however, leave additional imprints in the light curve even if they do not transit. These are the so-called phase curve variations that include ellipsoidal, reflection and beaming effects.}
   {In Millholland \& Laughlin (2017), the authors scrutinized the \textit{Kepler} database looking for these phase variations from non-transiting planets. They found 60 candidates whose signals were compatible with planetary companions. In this paper, we perform a ground-based follow-up of a sub-sample of these systems with the aim of confirming and characterizing these planets and thus validating the detection technique.}
   {We used the CAFE and HERMES instruments to monitor the radial velocity of ten non-transiting planet candidates along their orbits. We additionally used AstraLux to obtain high-resolution images of some of these candidates to discard blended binaries that contaminate the \textit{Kepler} light curves by mimicking planetary signals.}
   {Among the ten systems, we confirm three new hot-Jupiters (KIC\,8121913\,b, KIC\,10068024\,b, and KIC\,5479689\,b) with masses in the range 0.5-2 M$_{\rm Jup}$ and set mass constraints within the planetary regime for the other three candidates (KIC\,8026887\,b, KIC\,5878307\,b, and KIC\,11362225\,b), thus strongly suggestive of their planetary nature.}
   {For the first time, we validate the technique of detecting non-transiting planets via their phase curve variations. We present the new planetary systems and their properties. We find good agreement between the RV-derived masses and the photometric masses in all cases except KIC\,8121913\,b, which shows a significantly lower mass derived from the ellipsoidal modulations than from beaming and radial velocity data.}


   \maketitle
%
\section{Introduction}

In the era of high-precision photometric space-based missions, the detection of extrasolar planets through the transit technique has been very fruitful. So far, more than 4300 planets have been found around other stars, with around three fourths of them detected by this technique. Since the launch of the CoRoT mission \citep{auvergne09}, the subsequent NASA planet hunter \textit{Kepler} \citep{borucki10} and its extended \textit{K2} mission \citep{howell14} as well as its successor \textit{TESS} \citep{ricker14} have scrutinized the sky looking for these planetary fingerprints in the light curves of stellar sources.  According to the NASA Exoplanet archive, more than 2400 planets were discovered with \textit{Kepler}, more than 400 with \textit{K2}, and presently 137 have been detected and confirmed by \textit{TESS} using the transit method \citep{akeson13}.

Several years after the end of the prime \textit{Kepler} mission, the high-quality data provided by this telescope are still being profitably mined. Although \textit{Kepler} was tasked to find Earth-size planets around Solar-like stars, it explored an unexpectedly large range of planetary and orbital properties, including the first circumbinary planets, super-Earths, planets in highly eccentric orbits, and planets around giant stars, etc. (e.g., \citealt{doyle11,borucki12,ciceri14,lillo-box14}, respectively).

The transit technique, however, does not by itself provide a full confirmation of the planet candidate signal, as other non-planetary configurations can mimic the imprint of a symmetric dip in the light curve (e.g., background eclipsing binaries, grazing binary eclipses, etc.). Also, any blended source residing inside the photometer aperture can contaminate the light curve and provide inaccurate estimates of the planet candidate's parameters. Hence, additional follow-up is necessary from the ground. In practice, the confirmation of transiting planet candidates stems primarily from radial velocity (RV) follow-up, which can provide the mass of the transiting body and thus confirm its planetary nature.

\begin{figure}
\centering
\includegraphics[width=0.45\textwidth]{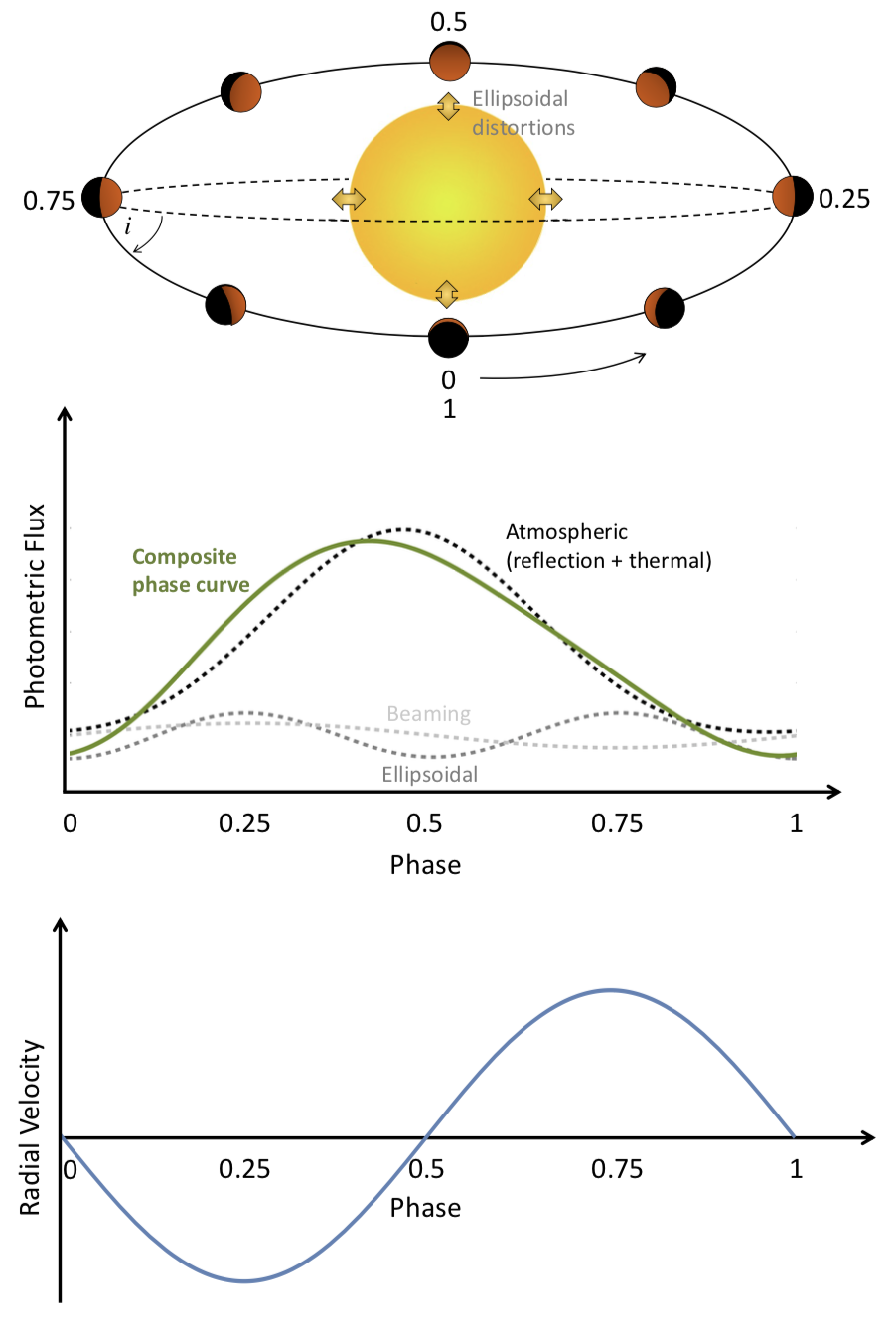}
\caption{Schematic representation of the reflection, ellipsoidal and beaming (REB) modulations. Top panel shows the orbital path of the planet at different phases and two orbital inclinations. Middle panel shows the corresponding shape of the REB effects (dotted labelled lines) and the composite phase curve (green solid line). The bottom panel shows the corresponding RV signal with its appropriate phase relationship.}
\label{fig:schematic}
\end{figure}

The extraordinary quality of the \textit{Kepler} photometry motivated the proposal of an alternate technique to detect and fully characterize (i.e., measure the mass and orbital configuration of) new planets. Previously used in binary systems, the detection of periodic modulations coupled with the orbital phase of the planet arose as a new potential detection technique (e.g., \citealt{faigler11}). These photometric variations are mainly due to three effects: reflected stellar light, ellipsoidal modulations, and Doppler beaming (hereafter the REB modulations, see Fig.~\ref{fig:schematic}). The reflection effect depends directly on the planet-to-star separation, size, and the geometric albedo of the planet. The ellipsoidal variations are known to be caused by tidal effects produced by a close-in object (e.g., a planet) on the external layers of the star (\citealt{morris85}). Since the amplitude of this modulation depends on the mass of the planet, it can potentially be used to both detect, confirm, and characterize planets. Finally, the Doppler beaming effect represents the photometric signature of the RV variations of the star induced by the planet motion when photometry is obtained on a fixed wavelength range \citep{Hills74,Rybicki79}. The induced RV implies that the stellar spectrum is periodically blue-shifted and red-shifted. As a consequence, more or less light enters in the wavelength domain of the photometric filter, yielding sinusoidal flux variations with an amplitude that depends on the mass of the planet. 

The REB modulations have already been used to detect and characterize close-in transiting planets from the \textit{Kepler} sample (e.g., \citealt{mazeh12,lillo-box14}). These systematic searches have established several interesting findings, including generally low (but occasionally quite large) hot Jupiter albedos \citep{demory11} and repeated detections of shifts in the maximum of the phase curve away from the sub-stellar point. Though it is substantial, the sample size ($\sim$~15) of \textit{Kepler} transiting planets with detectable photometric variations (less than 5 with RV confirmations) is short of the count necessary to adequately understand the atmospheric properties in a population sense. It is therefore worthwhile to increase the number of planets with well-characterized optical phase curves by expanding the search to non-transiting planets. Moreover, the masses derived from these modulations seem to not fully match the measurements from RVs in the few (less than 5) cases where both techniques could be applied. This is theoretically attributed to the presence of hot stream clouds in the atmosphere of the hot-Jupiters causing a phase shift in the light curve modulations.

\cite{millholland17} used a supervised machine learning approach to perform a systematic search of these modulations in the whole sample of 142\,630 FGK \textit{Kepler} stars without confirmed planets or KOIs (Kepler Objects of Interest). They identified 60 high-probability ($>$98\%) non-transiting hot Jupiter candidates that could readily be confirmed with RVs (see Fig.~1). In this paper, we present the ground-based follow-up of the brightest sub-sample of these candidates. In Sect.~\ref{sec:observations}, we present the targets and the observations. In Sect.~\ref{sec:analysis}, we perform a joint analysis of the light curve and RV data to confirm the planetary nature of some of the candidates and constrain the masses of the remaining sub-sample. In Sect.~\ref{sec:discussion}, we discuss the results of the analysis, and we conclude in Sect.~\ref{sec:conclusions}.

\section{Follow-up observations}
\label{sec:observations}

\subsection{Target selection}

This work is based on the follow-up of the \textit{Kepler} non-transiting planet candidates detected by \cite{millholland17} through the phase variations technique. Among the 60 candidates, we focused on those feasible with the Calar Alto Fiber-fed Echelle spectrograph (CAFE, \citealt{aceituno13, lillo-box20}). We selected ten of the brightest candidates (Kp $<14$ mag) displaying photometric periods of a few days ($P<5$ days) and with estimated RV amplitudes ($K\sin{i}$)\footnote{{Estimation of $K\sin{i}$ is non-traditional, but it arises directly from a measurement of the photometric amplitude of the ellipsoidal variation, $A_{\mathrm{ellip}}$. That is, $K\sin{i} \propto A_{\mathrm{ellip}} \propto M_p\sin^2{i}$. (See equation \ref{eq: Aellip} ahead or equation 27 of \cite{millholland17}.) This explains the extra factor of $\sin{i}$.} } from the photometric phase curve analysis feasible with CAFE ($K\sin{i}>10$~m/s). Table \ref{tab:targets} lists our initial selection of targets. 

\setlength{\extrarowheight}{2pt}
\begin{table}
\centering
\caption{List of targets selected for initial follow-up. The columns are the Kepler Input Catalog (KIC) target ID, the Kepler magnitude $K_p$ \citep{Mathur17}, and the estimated orbital period $P$ and $K\sin{i}$ of the candidate non-transiting hot Jupiter determined from \cite{millholland17} {(where $K$ is the radial velocity semi-amplitude and $i$ is the unknown orbital inclination)}, and whether they were observed in high-resolution imaging (HRI). The last four targets correspond to those with large rotation velocities preventing precise radial velocity extraction. }
\begin{tabular}{l | c c c c}
KIC & $K_p$ & $P$ (days) & $K\sin{i}$ (m/s) & HRI \\
\hline
8121913 & 11.7 & $3.2943\pm0.0003$ & $61\pm66$   & Yes \\
11362225 & 12.0 & $2.7513\pm0.0021$ & $103\pm71$   & Yes  \\ 
10068024 & 13.1 & $2.0735\pm0.0012$ & $229\pm243$   & Yes  \\
5878307 & 13.6 & $2.0466\pm0.0013$ & $116\pm112$   & Yes  \\
5479689 & 13.8 & $1.7012\pm0.0008$ & $158\pm111$   & Yes  \\
8026887 & 13.9 & $1.9232\pm0.0009$ & $234\pm113$   & Yes\tablefootmark{$\dagger$}  \\
6783562 & 13.6 & $2.0883\pm0.0014$ & $243\pm267$   & No  \\
5001685 & 14.0 & $2.2906\pm0.0014$ & $221\pm108$   & No  \\
8042004 & 13.3 & $1.9498\pm0.0011$ & $42\pm39$   & No  \\
10931452 & 12.7 & $2.7001\pm0.0023$ & $108\pm91$   & Yes  \\
\hline
\end{tabular}
\label{tab:targets}
\tablefoot{
\tablefoottext{$\dagger$}{The quality of this high-spatial resolution image is bad due to sudden worsening of the weather conditions.}}
\end{table}


\subsection{\textit{Kepler} light curves}
\label{sec: Kepler light curves}

In Section \ref{sec: joint radial velocity and light curve modeling}, we will perform a joint fit of the RV and light curve data. For the \textit{Kepler} light curves, we use the same pre-processing techniques as \cite{millholland17}; further details can be found therein. We began with the pre-search data conditioning (PDC) photometry \citep{smith12, stumpe12, Stumpe14} provided by the \textit{Kepler} Science Center. We performed two types of detrending: (1) an initial sliding polynomial fit, which was first used to identify candidate periodic signals using a Lomb-Scargle periodogram, and (2) a two-component detrending on the candidate signals, which simultaneously fit a sinusoidal phase curve component and a sliding polynomial stellar variability component. Using this latter fit, we created the detrended light curve by dividing the PDC light curve by the polynomial component of the fit. These detrended light curves will be used in Section \ref{sec: joint radial velocity and light curve modeling}, where we will phase fold and bin the light curve data for our joint fit.

\subsection{Astralux high-spatial resolution imaging}

We used the AstraLux instrument \citep{hormuth08} at the 2.2m telescope in Calar Alto observatory to obtain high-spatial resolution images of seven of the selected candidates. AstraLux is a simple fast-readout camera that performs the so-called lucky imaging technique on a 24x24 arcsec maximum field-of-view (FOV), with the possibility of windowing  to smaller FOVs in order to gain in readout time and thus reduce the exposure time. The instrument has demonstrated an excellent performance and has already shown a key contribution to the Kepler Follow-up Observing program (KFOP) in \cite{lillo-box12,lillo-box14,furlan17} and in key results from the K2 (e.g., \citealt{santerne18,barros17}) and TESS (e.g., \citealt{armstrong20,bluhm20,soto21}) missions.

\begin{figure*}
\centering

\includegraphics[width=0.4\textwidth]{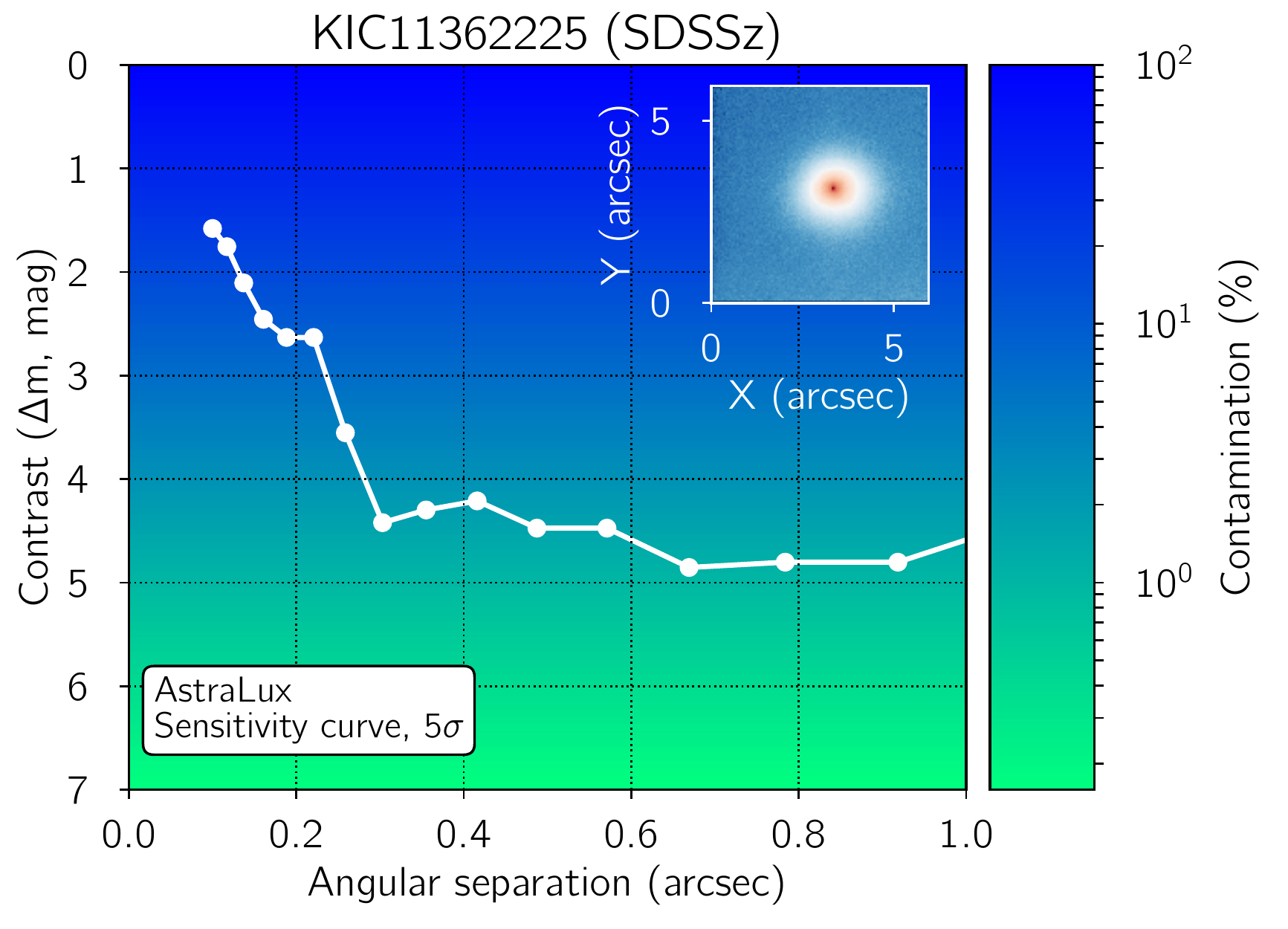}\includegraphics[width=0.4\textwidth]{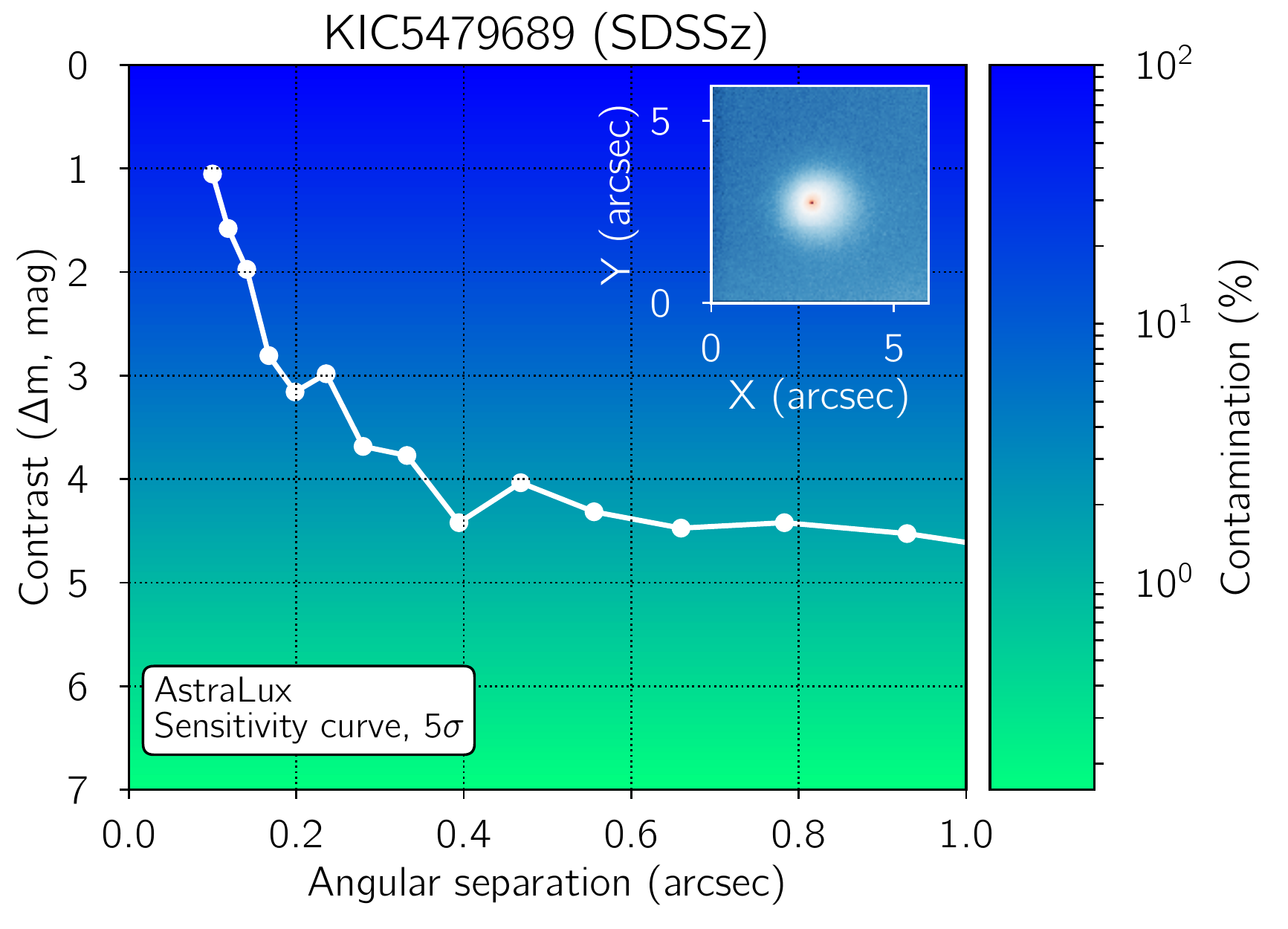}

\includegraphics[width=0.4\textwidth]{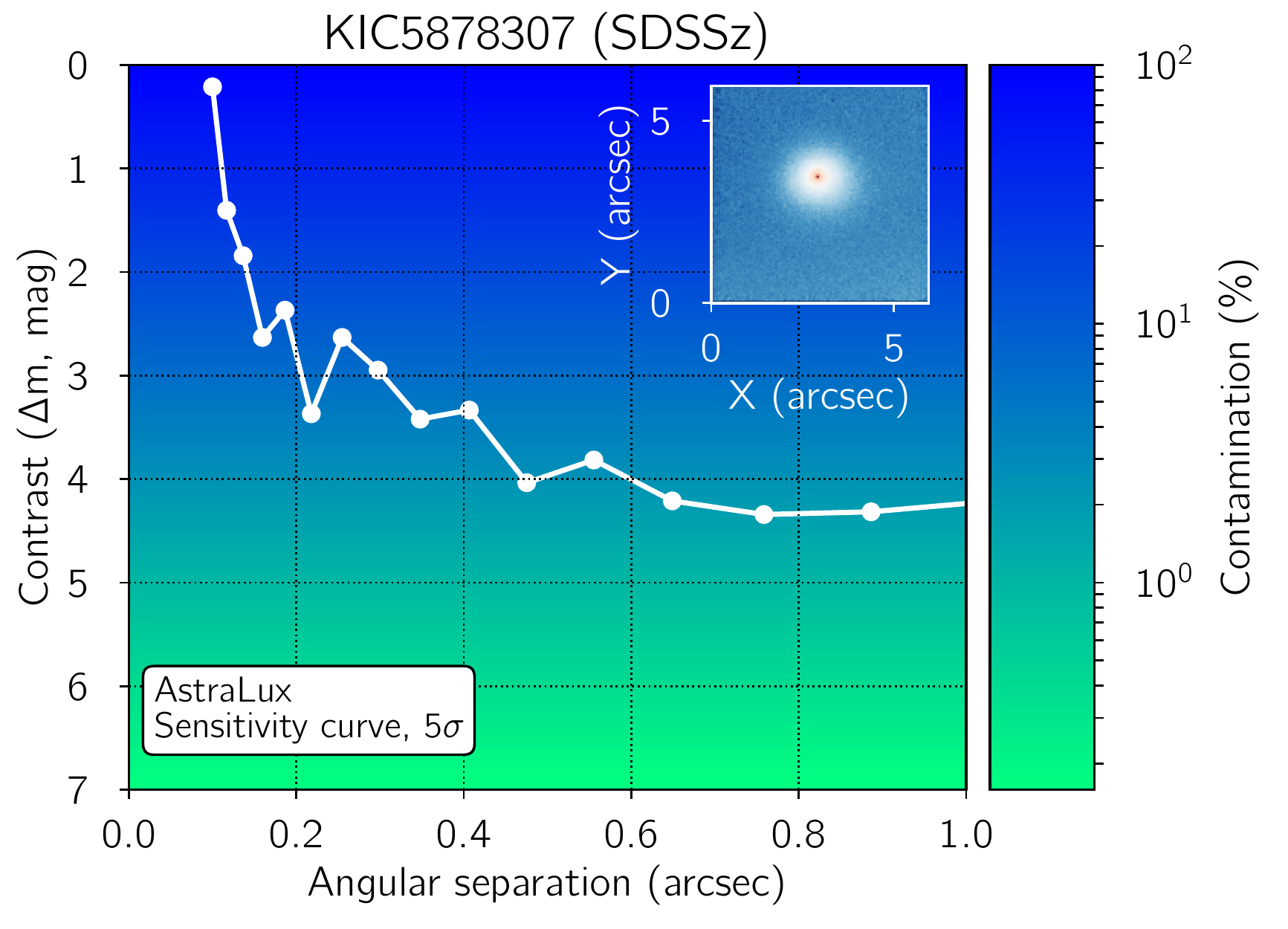}\includegraphics[width=0.4\textwidth]{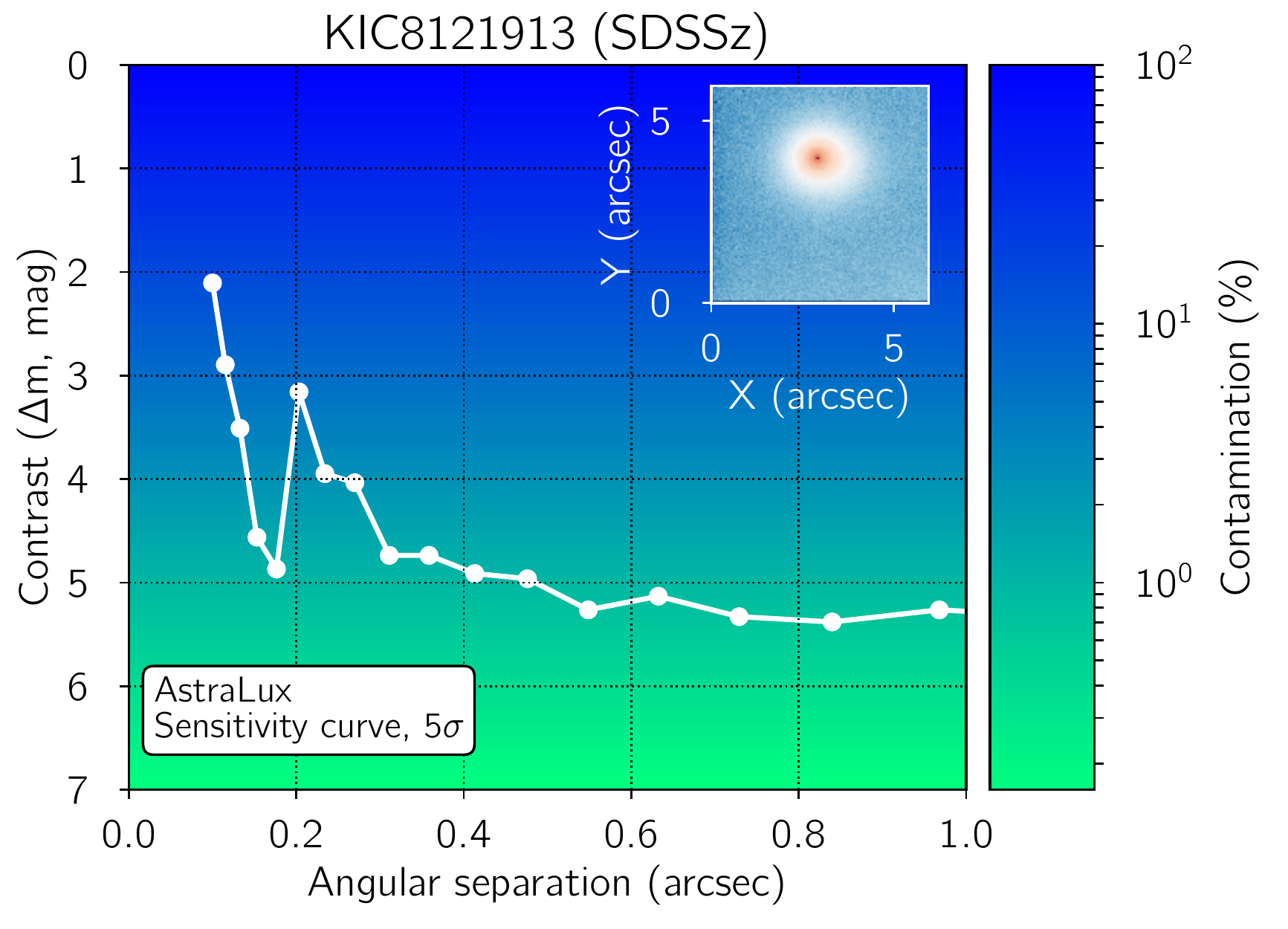}

\includegraphics[width=0.4\textwidth]{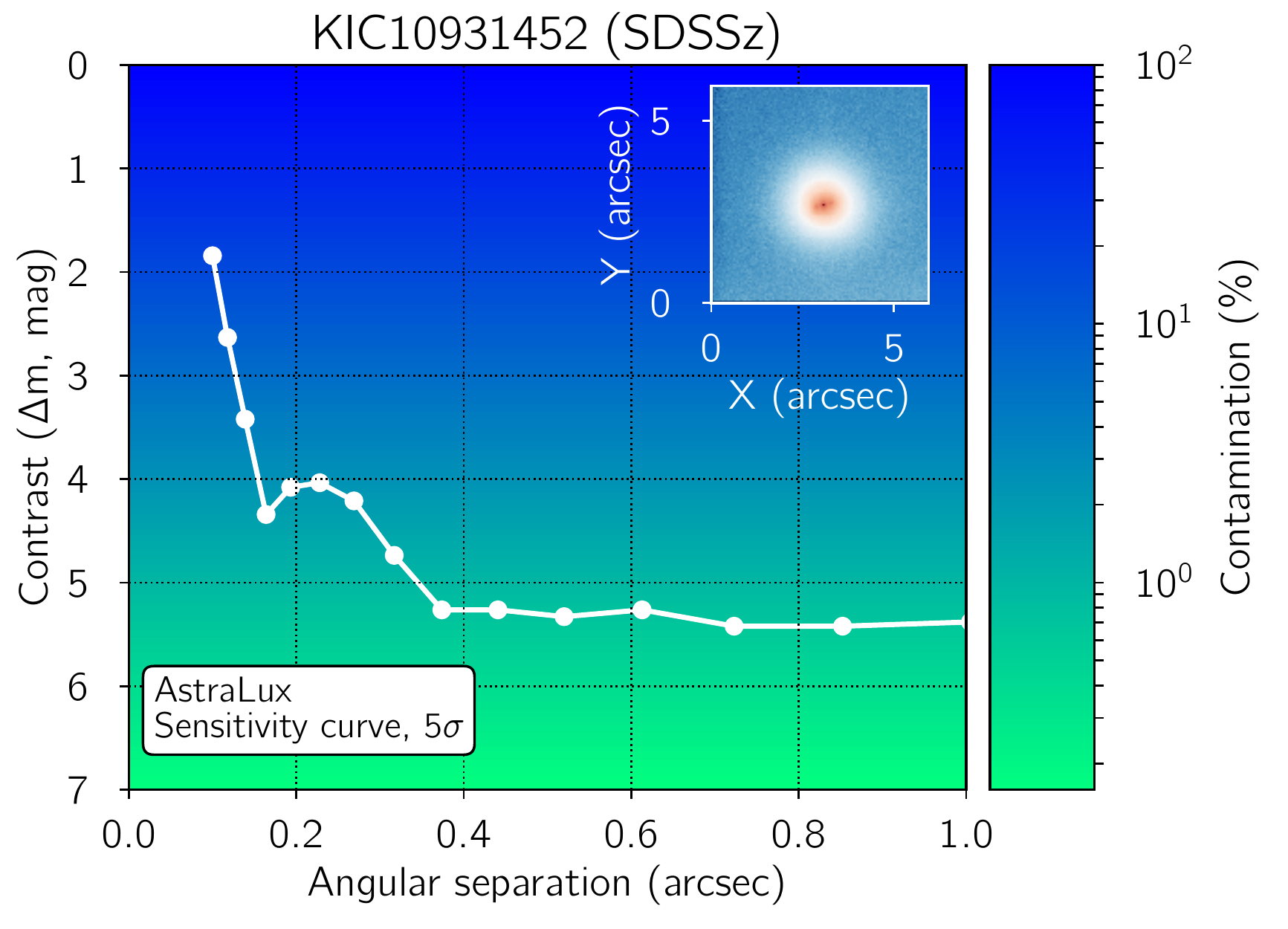}\includegraphics[width=0.4\textwidth]{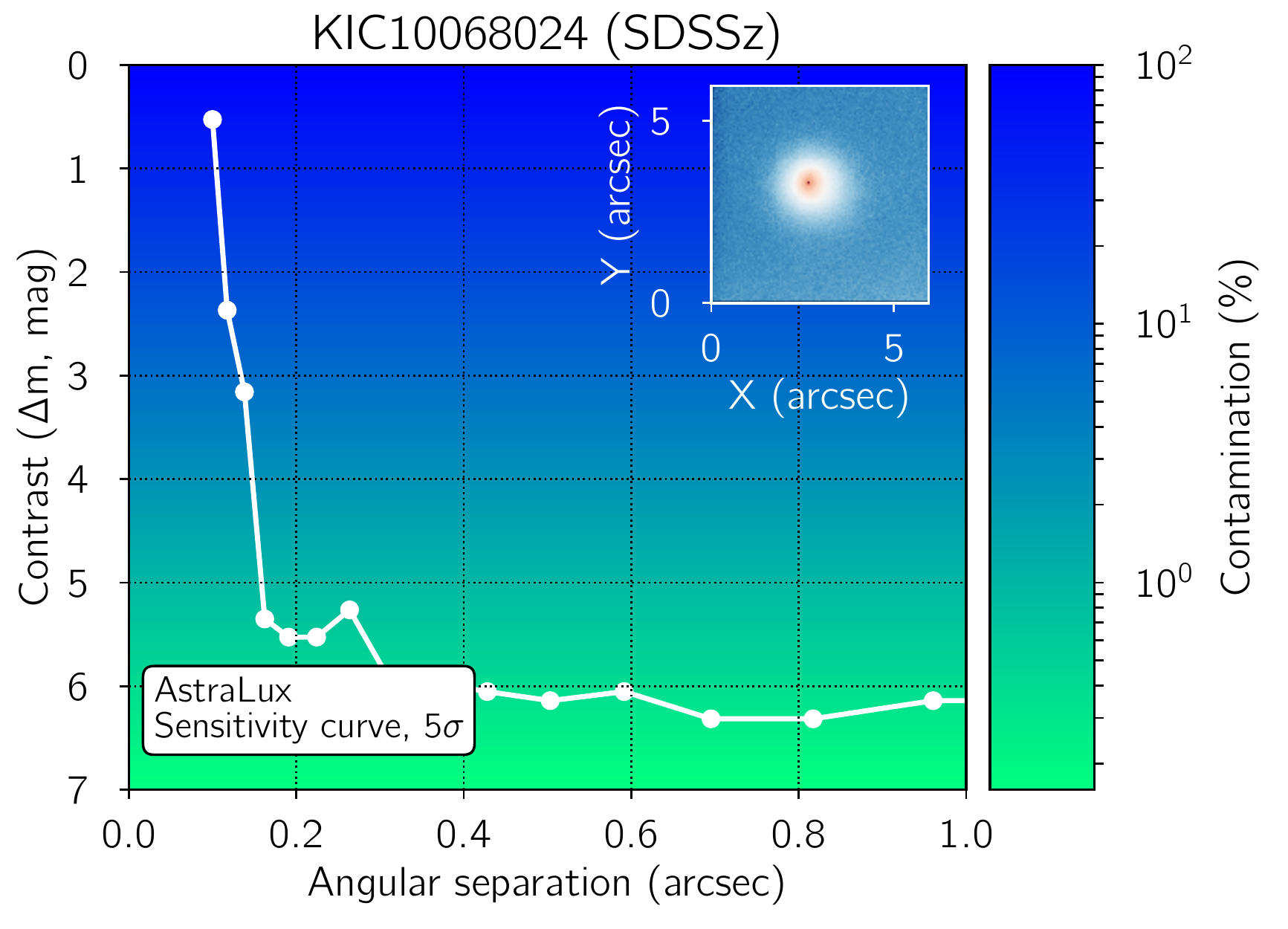}

\caption{Results of the analysis of the high-spatial resolution images obtained with the AstraLux lucky-imaging instrument. Each panel shows the contrast curve calculated as in \cite{lillo-box14} as a white line. The color-code of the contrast curve represents the contamination produced by a blended star of the given magnitude contrast as shown in the color bar. The inset panel displays the AstraLux image, with redder colors indicating more intensity.}

\label{fig:astralux}
\end{figure*}

Seven of the targets were observed with AstraLux on 13 August 2019 under 1.2 arcsec mean seeing conditions and good transparency conditions. One of the images (corresponding to KIC\,8026887) is too noisy to be useful given a sudden worsening of the weather conditions. Hence, we here analyze the remaining images for six of our targets. The datacubes were reduced with the automatic pipeline running at the observatory \citep{hormuth08}. This pipeline performs the basic reduction of the data and then selects the 10\% best frames (according to their Strehl ratio, \citealt{strehl1902}) and aligns and stack the frames to get the final high-spatial resolution image. These images are subsequently analyzed by our pipeline to detect possible blended sources and to obtain the contrast curves and estimate the maximum contamination in the \textit{Kepler} light curves (see \citealt{lillo-box12,lillo-box14} for details on how these contrast curves are estimated). The resulting contrast curves and AstraLux images are presented in Fig.~\ref{fig:astralux}.

The high-resolution images\footnote{All processed AstraLux images are publicly available through the Calar Alto Archive maintained by the Spanish Virtual Observatory at \url{http://caha.sdc.cab.inta-csic.es/calto/}} do not show any signs of close companions within 3 arcsec for any of the observed targets. Additionally, the contrast curves discard photometric contamination from undetected companions below 1\%. This is sufficient to ensure that the source of the photometric phase variations detected in \cite{millholland17} is the bright (and only) target in the field of view. Given this result, we will assume throughout our analysis that the only source in the \textit{Kepler} aperture and hence the only contributor to the photometric signal is the target star.

\subsection{CAFE high-resolution spectra}

We used the CAFE instrument \citep{aceituno13} with its recently upgraded capabilities \citep{lillo-box20} to perform a systematic RV follow-up of the selected planet candidates. CAFE is a fiber-fed high-resolution (R$=63\,000$) echelle spectrograph located in the coud\'e room of the 2.2m telescope at the Calar Alto observatory (Almer\'ia, Spain). Although the instrument has been operating since 2012, in 2016 its grating suffered a strong degradation that affected its efficiency and necessitated its retirement. This was taken as an opportunity to perform additional upgrades alongside the exchange of the grating. These improvements included a new double scrambler, introduction of temperature control of the isolated room where the instrument is located, and development of a new Python-based public pipeline (\textit{CAFExtractor}, see \citealt{lillo-box20}) that automatically reduces the data and provides high-level data-products like precise RV measurements using the cross-correlation technique \citep{baranne96}. These new capabilities were commissioned during 2018 and 2019, and the instrument was brought to full operation in April 2019. We refer the interested reader to \cite{lillo-box20} for additional details on the new capabilities of CAFE.

The observations were carried out in May-August 2019 under the approved program ID H19-2.2-007 (PI: Lillo-Box). In total, we obtained 53 spectra from the ten targets. We used the new \textit{CAFExtractor} pipeline to reduce the spectra and obtain the RVs. The pipeline includes an option to fit the CCF by using either a Gaussian profile (by default) or a rotational profile for the rapidly rotating stars (recommended for stars with v$\sin{i} > 10$~km/s). The rotational profile is defined as in \cite{gray05} and it has already been used in several programs that obtained precise RV measurements of rapidly rotating stars (e.g., \citealt{santerne12a,aller18}). For cases with v$\sin{i} > 10$~km/s, we used this approach, while for v$\sin{i} < 10$~km/s, we used the usual Gaussian profile. In four out of the ten observed targets, either the CCF was too broad to extract meaningful RVs (KIC\,6783562 and KIC\,10931452) or the phase coverage was insufficient to perform a relevant analysis (KIC\,5001685 and KIC\,8042004). Hence, these targets are not further addressed in this paper. The extracted RVs and the properties of the CCF for the remaining six targets are shown in Tables~\ref{tab:RV_start} to \ref{tab:RV_end}.

\subsection{HERMES/Mercator high-resolution spectra}
We observed five of the targets from the selected sample with the high-resolution fiber-fed HERMES spectrograph \citep{raskin11} at the Mercator telescope in La Palma in August 2020. HERMES is a sibling of the CORALIE instrument at La Silla. It provides a resolution of R=85\,000 and a wavelength coverage of 377-900\,nm. We used the simultaneous wavelength calibration mode, which uses a second fiber to inject a Thorium-Argon spectrum in between the science orders to monitor the nightly drift of the wavelength calibration for precise RV measurements. The instrument's automatic pipeline provides the basic reduction of the spectra. We measured the RVs by following a similar procedure as described for the CAFE spectra, with a new mask adapted for this wavelength coverage and resolution. The extracted RVs and the properties of the CCF are shown in Tables~\ref{tab:RV_start} to \ref{tab:RV_end}.

\section{Analysis}
\label{sec:analysis}

\subsection{Radial velocity standalone modeling}
\label{sec:cafe}
We first performed a RV analysis of the CAFE data, since the number of RV measurements from the HERMES instrument was low for this preliminary analysis. Given the limited number of observations for most of the systems, we used a simplistic RV model that assumes circular orbits (likely a good approximation given the short periods of these candidates) and adopts Gaussian priors on the period and time of conjunction obtained from the light curve analysis in \cite{millholland17}. Consequently, the only two free parameters of the model are the systemic velocity (V$_{\rm sys}$) and the semi-amplitude of the RV signal ($K$). For each of these parameters, we assumed a uniform prior with broad ranges corresponding to $\mathcal{U}(-100,100)$~km/s for V$_{\rm sys}$ and $\mathcal{U}(0.0, 10)$~km/s for $K$.

\begin{figure*}
\centering
\includegraphics[width=0.33\textwidth]{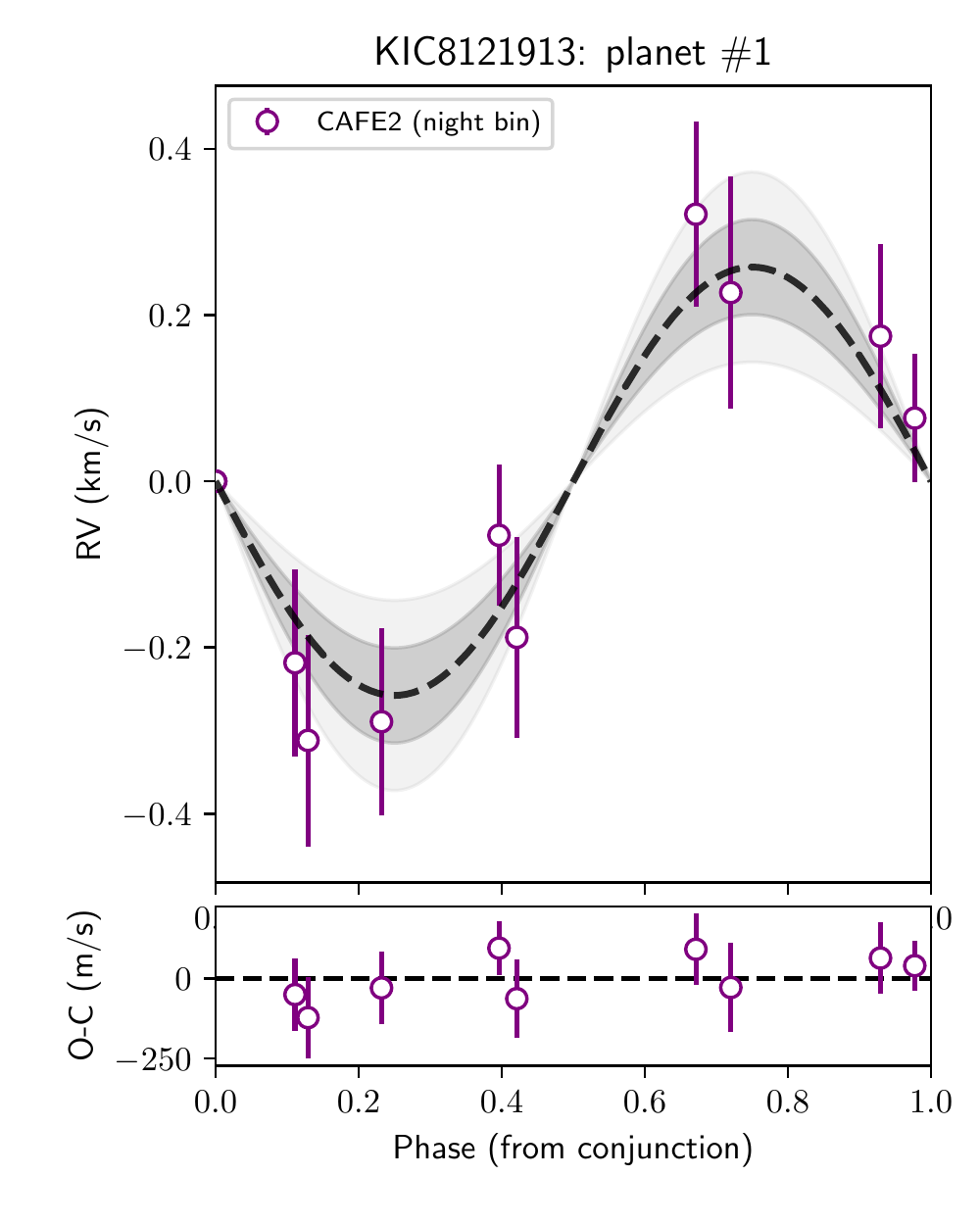}\includegraphics[width=0.33\textwidth]{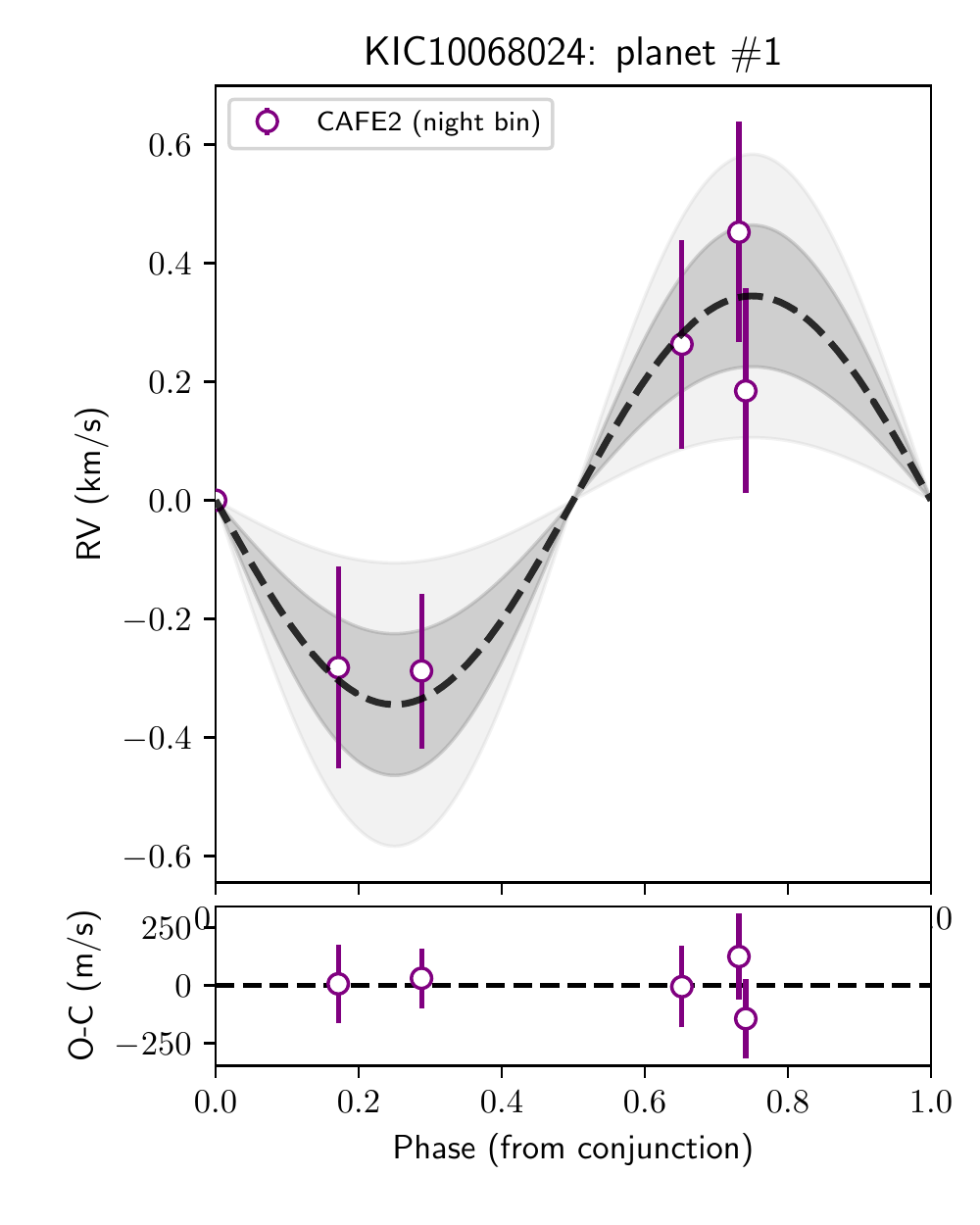}\includegraphics[width=0.33\textwidth]{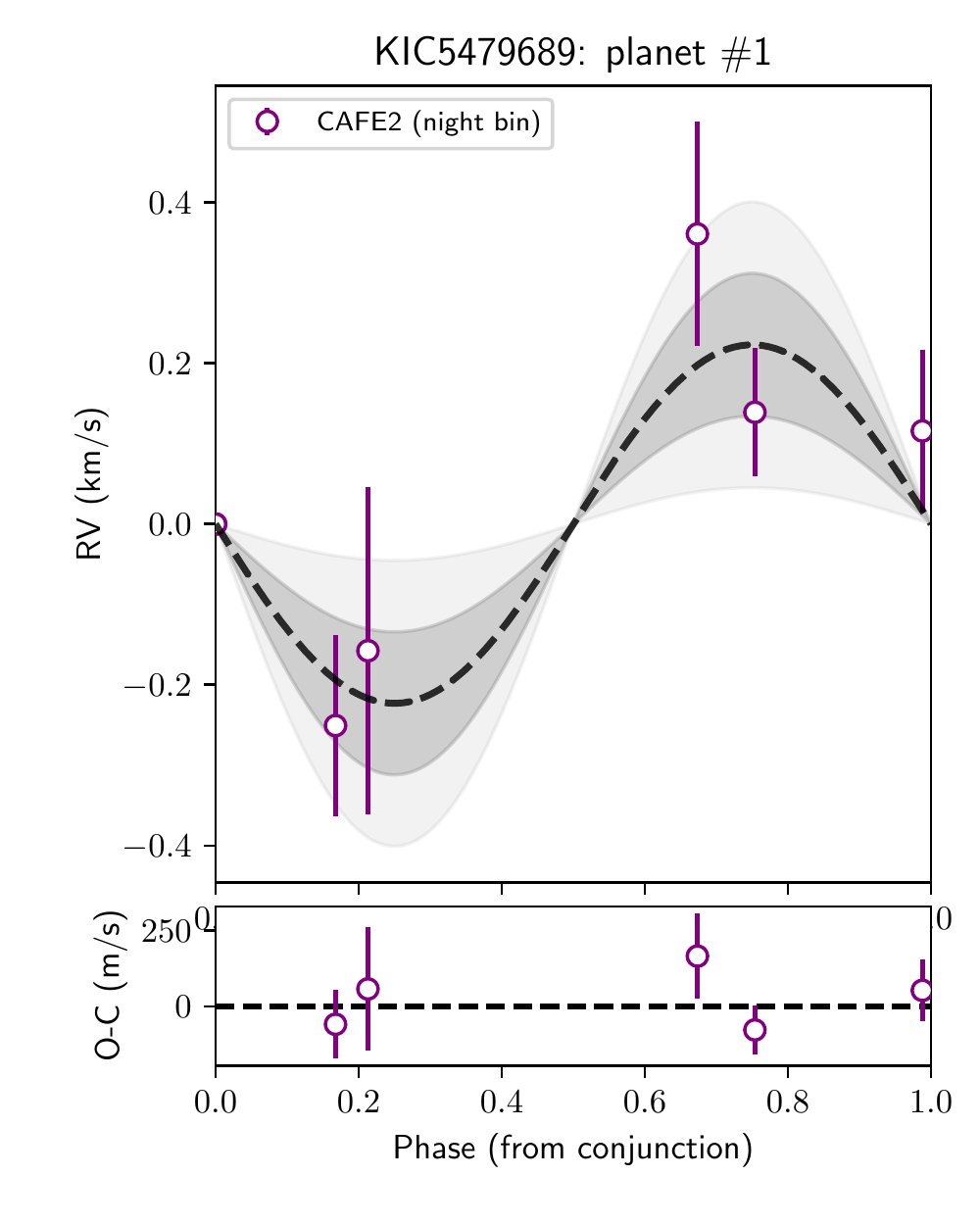}
\caption{Phase-folded radial velocity curves for the three planets confirmed in this paper. {The solid open symbols represent nightly binned values when more than one observation are available on a given night (although we note that the analysis in the paper has been done using the individual measurements). The median radial velocity model is represented by the black dashed line and the 67.8\% and 95\% confidence intervals are displayed as dark grey and light grey shaded regions around the median model.}}
\label{fig:rv1}
\end{figure*}

\begin{figure*}
\centering
\includegraphics[width=0.33\textwidth]{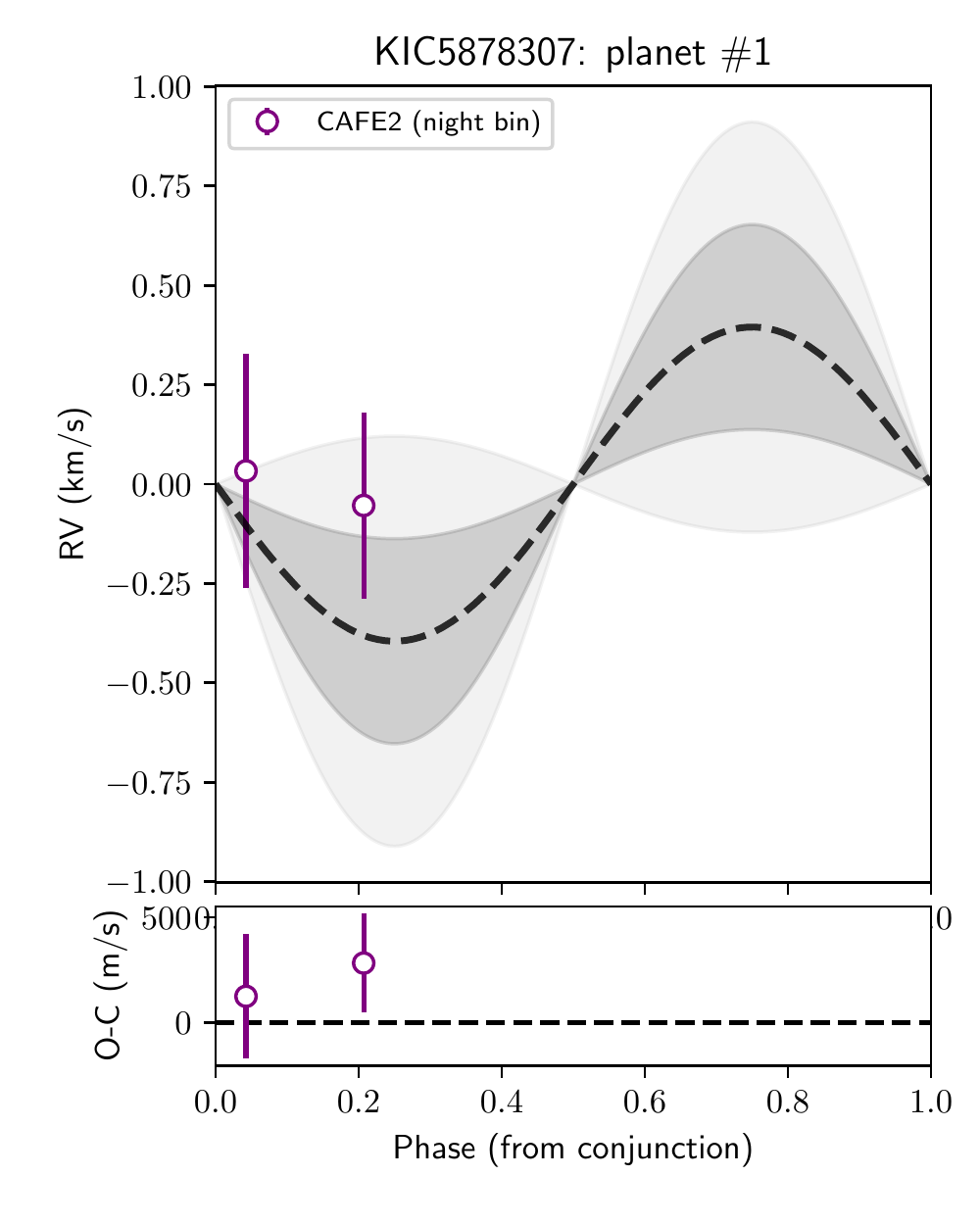}\includegraphics[width=0.33\textwidth]{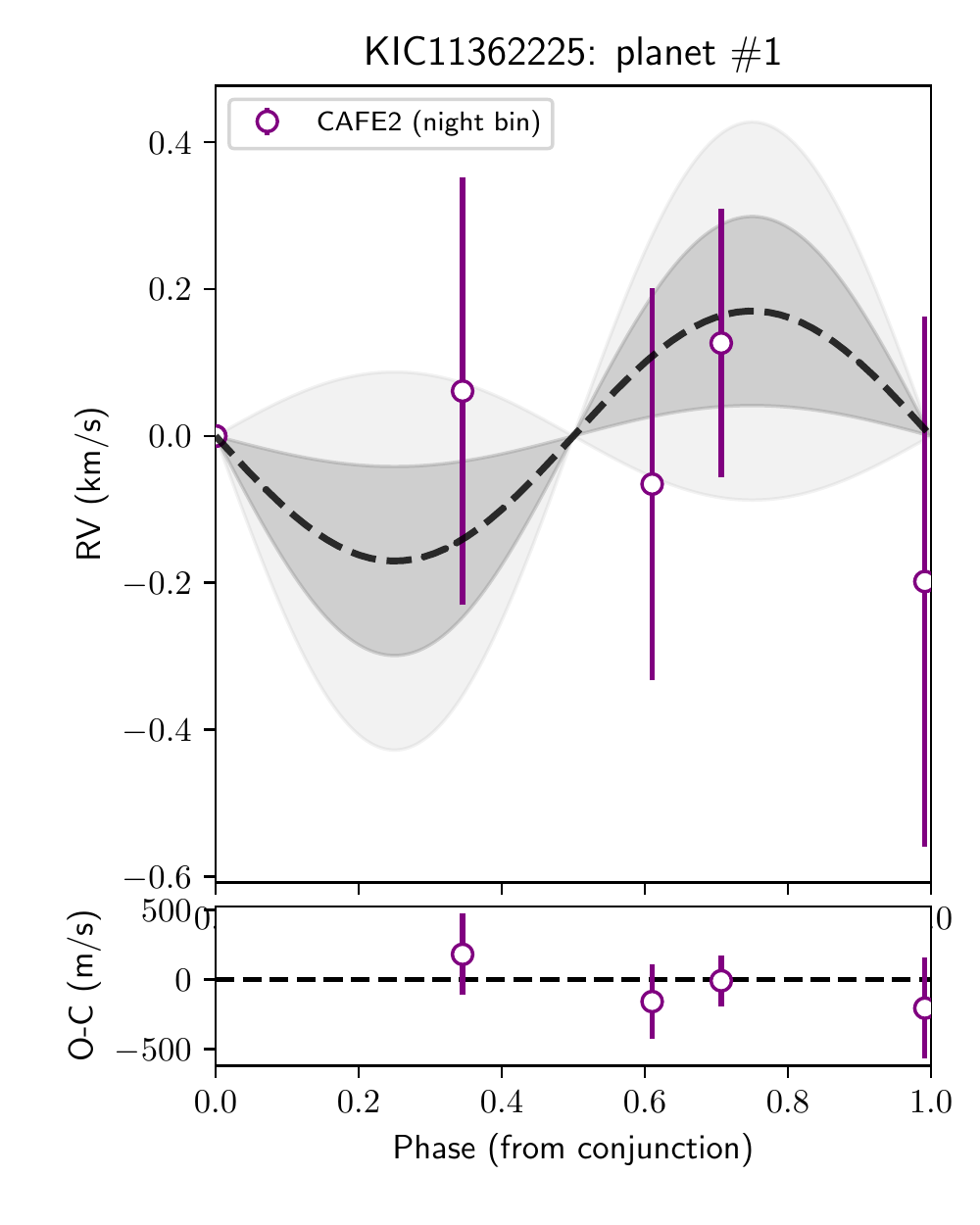}\includegraphics[width=0.33\textwidth]{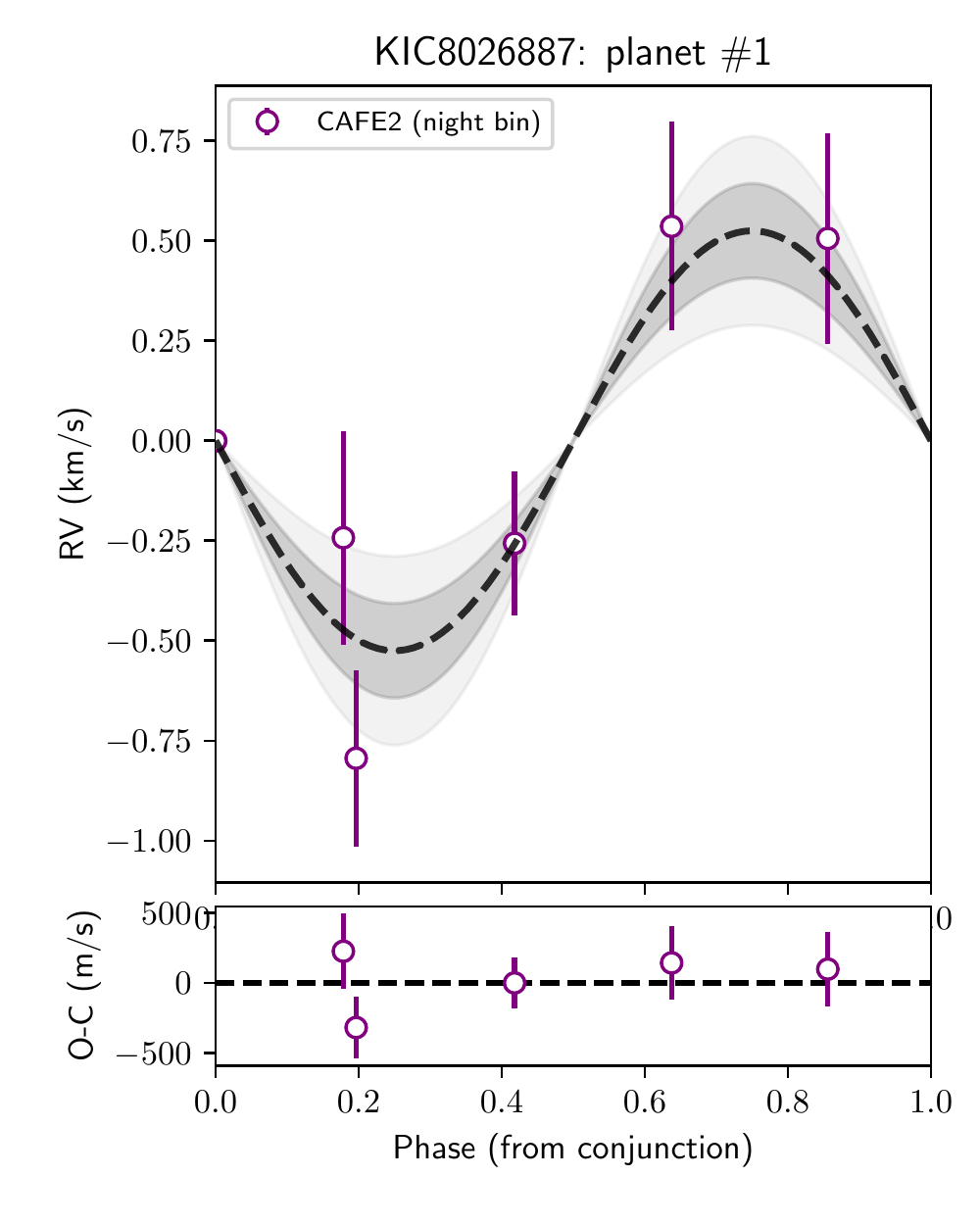}
\caption{Phase-folded radial velocity curves for the two planets where only upper limits to the planet mass are possible (left and middle panels) and the case of KIC 8026887 with the period equal to twice the photometric period found in \cite{millholland17}. {The solid open symbols represent nightly binned values when more than one observation are available on a given night (although we note that the analysis in the paper has been done using the individual measurements). The median radial velocity model is represented by the black dashed line and the 67.8\% and 95\% confidence intervals are displayed as dark grey and light grey shaded regions around the median model.}}
\label{fig:rv2}
\end{figure*}

We used the Markov Chain Monte Carlo affine invariant ensemble sampler \texttt{emcee} \citep{2010CAMCS...5...65G, 2013PASP..125..306F} to explore the posterior probability of the two parameters. We ran a first set of 20 walkers with 10\,000 steps per walker to identify the maximum likelihood location of the parameter space. Then we ran another 10 walkers with 10\,000 steps per walker in a nearby region around this maximum likelihood location of the parameter space to properly sample the posterior distributions. The results are shown in Figs.~\ref{fig:rv1} and \ref{fig:rv2}, and the priors, median and confidence levels of the posterior distributions for each parameter are shown in Tables~\ref{tab:RVposterior1} to \ref{tab:RVposterior2}. 

Among the six systems with appropriate phase coverage, we find clear detection in three of the targets (KIC\,8121913, KIC\,10068024, and KIC\,5479689), see Fig.~\ref{fig:rv1}. In these cases, the RV semi-amplitude is detected with a confidence larger than 99.7\% (corresponding to $3\sigma$) and thus are considered as confirmed planets. In two other cases (KIC\,5878307 and KIC\,11362225, see Fig.~\ref{fig:rv2}), the data is insufficient to measure the planet mass but the low amplitude of the RV variations and the absence of contaminants in the AstraLux images validate their planetary nature. {In the case of KIC\,8026887, we find either a large amplitude of several hundred of meters per second or a strong linear trend with a smaller RV amplitude. Additionally, by using a period equal to twice the photometric period ($2\times P=3.8464$ days), we find a clear signal compatible with a Keplerian model with $K\sim610\pm180$~m/s (see Fig.~\ref{fig:rv2}, right panel). This is further investigated in the joint light curve and radial velocity analysis in the next section.}



\subsection{Joint radial velocity and light curve modeling}
\label{sec: joint radial velocity and light curve modeling}

To examine the consistency of the RV observations and \textit{Kepler} light curves, we performed a simultaneous fit of the two datasets for the four targets with sufficiently precise RVs: KIC 8121913, KIC 10068024, KIC 5479689, and KIC 8026887. The key goal of this simultaneous fit is to verify that the periods and phases suggested by the photometric detections are consistent with the RVs. Moreover, we also wish to investigate whether the companion mass estimates agree between the two datasets. Here we adopt a fitting approach that is as simplified as possible, on account of our relatively small number of RV data points and our goal to simply demonstrate RV and light curve consistency. 

We model the phase curve variations using a four-component sum of sinusoids at the orbital phase 
\begin{equation}
\phi\equiv(t - T_0)/P,
\label{eq: phi}
\end{equation}
where $T_0$ is the epoch of inferior conjunction (or the transit time if the companion is transiting):
\begin{equation}
\begin{split}
F(\phi) &= A_0 - A_{\mathrm{refl}}\cos(2\pi(\phi + \delta)) \\
&- A_{\mathrm{ellip}}\cos(4\pi\phi) + A_{\mathrm{beam}}\sin(2\pi\phi).
\label{eq: F(phi)}
\end{split}
\end{equation}
Here, {$\delta$ is a phase offset in peak brigtness of the reflection component,} and $A_{\mathrm{refl}}$, $A_{\mathrm{ellip}}$, and $A_{\mathrm{beam}}$ are positive quantities representing the photometric amplitudes of the atmospheric/reflection, ellipsoidal, and beaming components of the phase curve \citep{esteves14, Shporer17}. These amplitudes will be treated as free parameters in our model. In Section \ref{sec: phase curve amplitudes}, we will describe the origin of these phase curve components and their approximate relationships to fundamental physical and orbital parameters. We will also verify that they have been fit with reasonable values.

This four-component phase curve model is fit to the detrended \textit{Kepler} light curves introduced in Section \ref{sec: Kepler light curves}. These light curves are phase-folded using the period $P$ and epoch $T_0$ from equation \ref{eq: phi} (both of which are free parameters in the fit) and binned into 400 bins, with the $j$th bin having mean flux, $F_j$, and standard deviation, $\sigma_{F,j}$. 

We now proceed to the RV component of the joint fit. We model the Doppler velocities simply as 
\begin{equation}
v(t) = -K\sin(2\pi\phi) + \dot{\gamma}(t-T_0) + \gamma_{\mathrm{inst}},
\label{eq: v(t)}
\end{equation}
where we have assumed a negligible orbital eccentricity. The RV semi-amplitude, $K$, is strictly positive, and $\phi$ and $T_0$ are the same phase and epoch as defined in equation~\ref{eq: phi}. In addition to the sinusoidal signal, we include a long-term linear trend with acceleration $\dot{\gamma}$ and instrument-specific offset, $\gamma_{\mathrm{inst}}$. 

We perform a joint RV and phase curve fit to the observations using the standard practices of Bayesian inference. The log-likelihood is here composed of the sum of the individual log-likelihoods of the RV and light curve datasets:
\begin{equation}
\ln\mathcal{L} = \ln\mathcal{L}_{\mathrm{RV}} + \ln\mathcal{L}_{\mathrm{LC}}.
\label{eq: lnL_tot}
\end{equation}
Assuming Gaussian-distributed noise, the RV log-likelihood is
\begin{equation}
\ln\mathcal{L}_{\mathrm{RV}} = -\frac{1}{2}\sum_j\left[{\frac{(v_j - v_{\mathrm{mod}}(t_j))^2}{(\sigma_{v,j}^2 + \sigma_{\mathrm{jit, inst}}^2)}} + \ln(2\pi[\sigma_{v,j}^2 + \sigma_{\mathrm{jit, inst}}^2])\right].
\label{eq: lnL_RV}
\end{equation}
Here, $v_j$ and $\sigma_{v,j}$ are the $j$th velocity measurement and measurement error, and $v_{\mathrm{mod}}(t_j)$ is the model RV (equation \ref{eq: v(t)}) at time $t_j$. 

Similarly, the light curve log-likelihood is given by
\begin{equation}
\ln\mathcal{L}_{\mathrm{LC}}= -\frac{1}{2}\sum_j{\left[\frac{(F_j - F_{\mathrm{mod}}(\phi_j))^2}{\sigma_{F,j}^2} + \ln(2\pi\sigma_{F,j}^2)\right]},
\label{eq: lnL_LC}
\end{equation}
where $F_j$ and $\sigma_{F,j}$ are the $j$th measurement and error resulting from the folded and binned light curve data. $F_{\mathrm{mod}}(\phi_j)$ is the model phase curve (equation \ref{eq: F(phi)}) at phase $\phi_j$.

The posterior distributions of the parameters were sampled through the Markov Chain Monte Carlo affine invariant ensemble sampler \texttt{emcee} \citep{2010CAMCS...5...65G, 2013PASP..125..306F} using uniform priors and the likelihood function given by equations \ref{eq: lnL_tot} --  \ref{eq: lnL_LC}. We collect 20,000 samples across 200 chains, discarding the first 5,000 samples as burn-in. Convergence was assessed by inspecting trace plots to see that the chains were well-mixed. We provide corner plots of the joint posterior parameter distributions in Appendix \ref{sec: posterior distributions}.

\begin{figure*}
\centering
\includegraphics[width=0.33\textwidth]{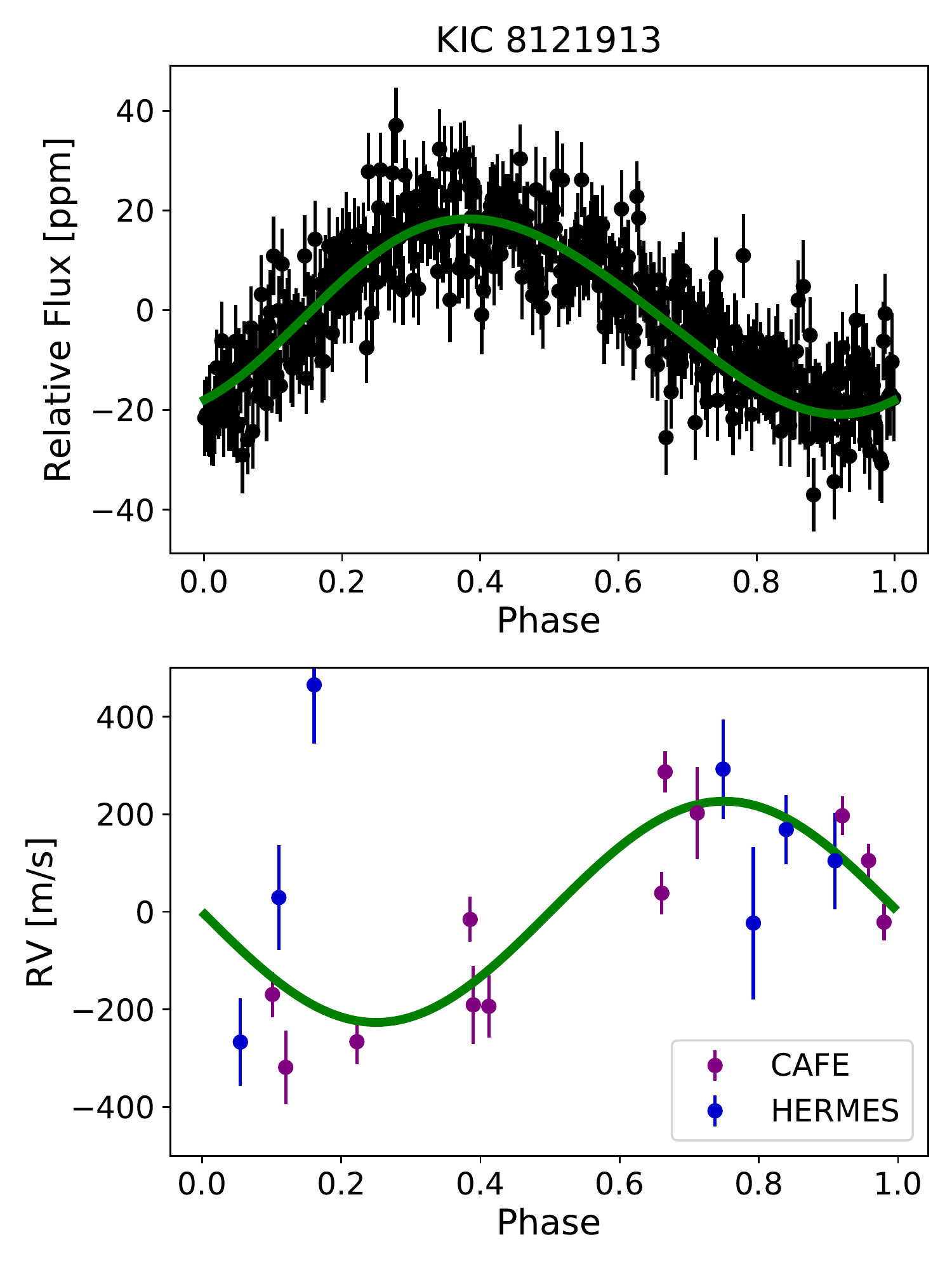}
\includegraphics[width=0.33\textwidth]{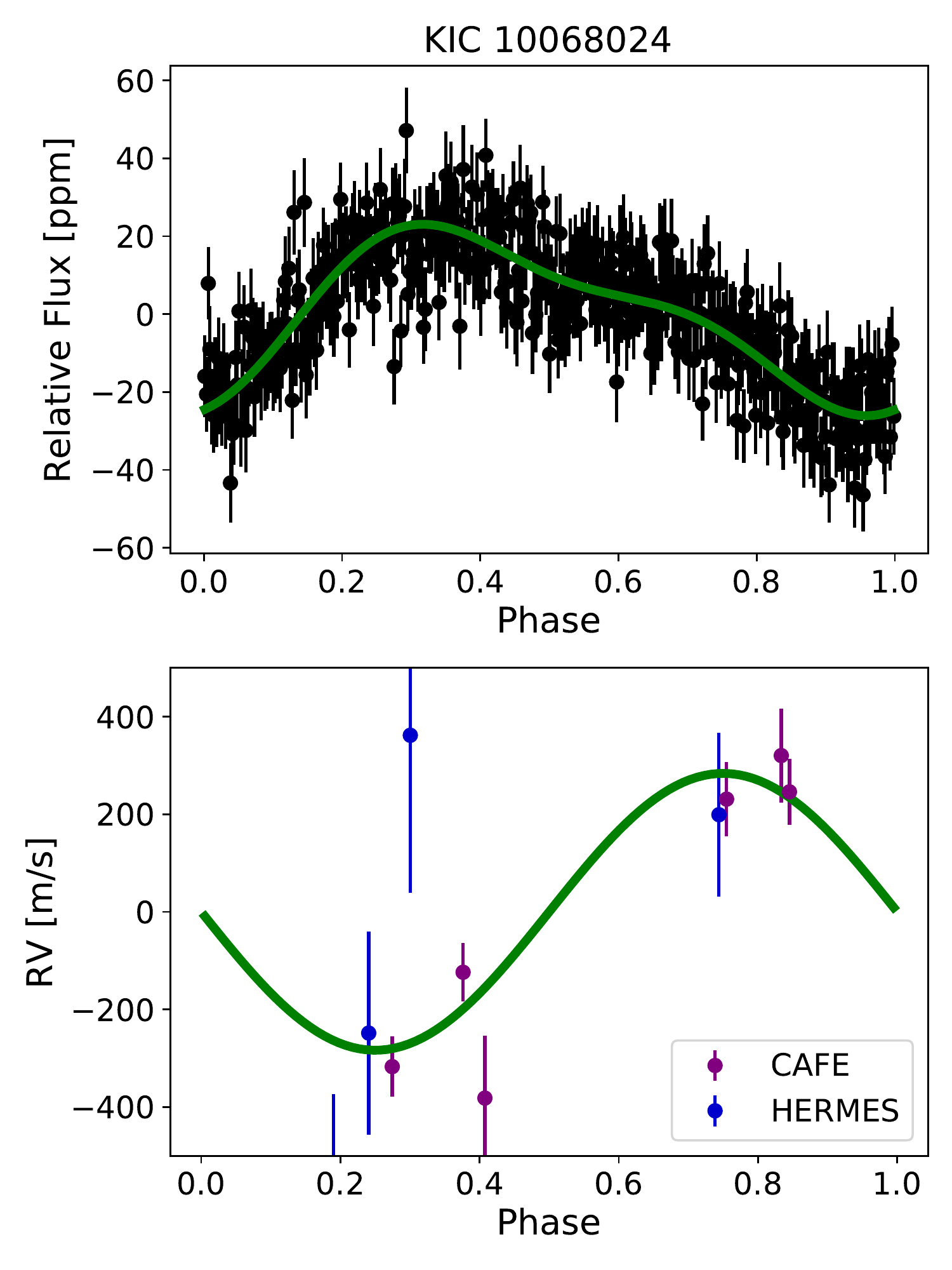}
\includegraphics[width=0.33\textwidth]{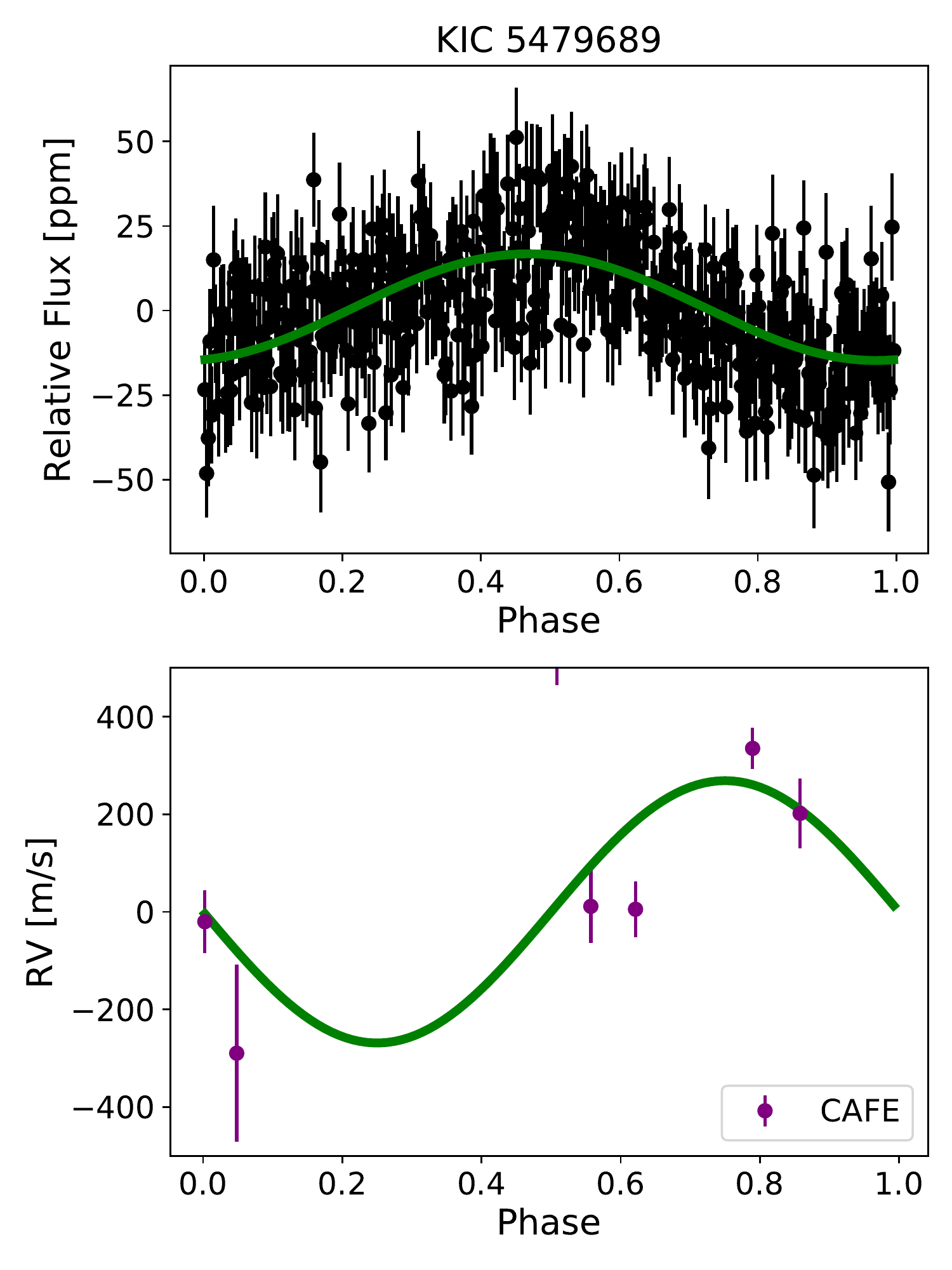}
\caption{Results of the joint phase curve (upper panels) and radial velocity (lower panels) fits for KIC 8121913, KIC 10068024, and KIC 5479689 (see Sect.~\ref{sec: joint radial velocity and light curve modeling})}
\label{fig: RV/phase curve fits}
\end{figure*}

The results of the joint fit are presented in Table~\ref{tab: joint fit}, and the best-fitting models for KIC 8121913, KIC 10068024, and KIC 5479689 are displayed in Figure \ref{fig: RV/phase curve fits}. We observe that consistent solutions for the joint RV/phase curve fit can be found for these systems. Specifically, there are sets of $P$ and $T_0$ for which the two datasets agree within uncertainties. The amplitudes of the phase curve and RV signals are also generally consistent in terms of the physical system parameters they correspond to, as we will discuss in Section \ref{sec: phase curve amplitudes}.

{In the case of KIC\,8026887, the phase curve clearly shows ellipsoidal variations when folded at $P = 1.9232$ days. Hence the double period hypothesis is unlikely. In any case, we tested it to ensure that the $P=1.9$ days is able to jointly model the RV and LC data. We then tested four models including priors on the period and turning on and off the linear trend. We then estimated the Bayesian evidence for the four models and concluded that the model with $P=1.9$ and the linear trend shows the largest evidence for the current dataset. This model is able to reconcile both RVs and light curve modulations and provides a planet mass of $3.44\pm0.42$~M$_{\rm jup}$. The difference in log-evidence is $+7$ compared to the model without a linear trend and $+135$ compared to the model with $2\times P$. The candidate absolute mass is hence significantly detected and we can hence validate the planetary nature of this candidate. Given the poor radial velocity phase coverage, however, we remain conservative and do not claim confirmation yet.}


\begin{table*}
\centering
\setlength{\extrarowheight}{4pt}
\caption{Results of the joint fits to the radial velocity and light curve data.}
\begin{tabular}{llll}
\hline
\hline
\label{tab: joint fit}
& KIC 8121913 & KIC 10068024 & KIC 5479689 \\
\hline
Orbital parameters \\
$T_0$ (BJD) &  $2456427.84 ^{+ 0.04 }_{- 0.05 }$ & $2456424.23 ^{+ 0.05 }_{- 0.05 }$  & $2456423.76 ^{+ 0.09 }_{- 0.15 }$ \\
$P$ (days)  &  $3.294601 ^{+ 0.000013}_{- 0.00022}$ & $2.073549 ^{+ 0.000008 }_{- 0.000177 }$  & $1.701531 ^{+ 0.000008 }_{- 0.000419 }$ \\
Phase curve parameters \\
$A_0$ (ppm)  & $-0.03 ^{+ 0.38 }_{- 0.38 }$  & $-0.08 ^{+ 0.52 }_{- 0.53 }$  & $-0.33 ^{+ 0.73 }_{- 0.81 }$ \\
$A_{\mathrm{refl}}$ (ppm) & $14.1 ^{+ 1.3 }_{- 1.1 }$ & $18.7 ^{+ 1.1 }_{- 1.0 }$  &  $14.1 ^{+ 1.3 }_{- 1.5 }$ \\
$A_{\mathrm{ellip}}$ (ppm)  & $2.2 ^{+ 0.5 }_{- 0.5 }$ & $5.8 ^{+ 0.8 }_{- 0.8 }$  & $0.3 ^{+ 0.5 }_{- 0.2 }$ \\
$A_{\mathrm{beam}}$ (ppm) & $11.8 ^{+ 2.8 }_{- 2.7 }$ &  $9.7 ^{+ 4.2 }_{- 4.3 }$ & $2.5 ^{+ 2.4 }_{- 1.8 }$ \\
$\delta$ &  $0.01 ^{+ 0.03 }_{- 0.03 }$ & $-0.0 ^{+ 0.04 }_{- 0.04 }$  & $0.02 ^{+ 0.02 }_{- 0.02 }$ \\
RV parameters \\
$K$ (m/s)  &  $224.14 ^{+ 36.95 }_{- 38.85 }$ & $276.7 ^{+ 46.4 }_{- 47.13 }$  & $82.37 ^{+ 65.24 }_{- 24.51 }$ \\
$\gamma_{\mathrm{CAFE}}$ (m/s)  &  $2826.4 ^{+ 2471.1 }_{- 2435.1 }$ &  $8582.9 ^{+ 4956.7 }_{- 4235.6 }$ & $26675.9 ^{+ 5116.7 }_{- 4930.2 }$ \\
$\gamma_{\mathrm{HERMES}}$ (m/s)  & $3278.6 ^{+ 2929.4 }_{- 2877.1 }$ & $9982.8 ^{+ 5759.0 }_{- 4933.8 }$  & $34211.5 ^{+ 10575.7 }_{- 10289.3 }$ \\
$\dot{\gamma}$ (m/s/day)  &  $-1.28 ^{+ 1.1 }_{- 1.12 }$ & $-3.8 ^{+ 1.88 }_{- 2.2 }$  & $-11.79 ^{+ 2.19 }_{- 2.28 }$ \\
$\sigma_{\mathrm{jit},\mathrm{CAFE}}$ (m/s)  &  $82.54 ^{+ 11.77 }_{- 15.54 }$ & $60.66 ^{+ 27.23 }_{- 36.66 }$  & $93.17 ^{+ 5.01 }_{- 10.47 }$ \\
$\sigma_{\mathrm{jit},\mathrm{HERMES}}$ (m/s)  &  $94.21 ^{+ 4.32 }_{- 9.02 }$ & $48.89 ^{+ 35.09 }_{- 33.5 }$  & $51.12 ^{+ 33.35 }_{- 34.3 }$ \\
Derived parameters \\
$M_p\sin{i}$ from $K$ ($M_{\mathrm{Jup}}$) &  $2.1 ^{+ 0.4 }_{- 0.4 }$ &  $2.0 ^{+ 0.4 }_{- 0.4 }$ & $0.5 ^{+ 0.4 }_{- 0.1 }$ \\
$M_p\sin{i}$ from $A_{\mathrm{ellip}}$ ($M_{\mathrm{Jup}}$) &  $0.3 ^{+ 0.2 }_{- 0.1 }$ & $0.8 ^{+ 2.1 }_{- 0.5 }$  & $0.1 ^{+ 0.2 }_{- 0.1 }$ \\
$M_p\sin{i}$ from $A_{\mathrm{beam}}$ ($M_{\mathrm{Jup}}$) &  $2.3 ^{+ 0.6 }_{- 0.5 }$ &  $1.5 ^{+ 0.6 }_{- 0.7 }$ & $0.3 ^{+ 0.3 }_{- 0.2 }$ \\
$\alpha_{\mathrm{ref}}{R_p}^2\sin{i} \ ({R_{\mathrm{Jup}}}^2)$ & $1.56 ^{+ 0.18 }_{- 0.15 }$ & $0.99 ^{+ 0.11 }_{- 0.11 }$  & $0.49 ^{+ 0.05 }_{- 0.05 }$ \\ \\
\hline
\hline
\end{tabular}
\end{table*}

\section{Discussion}
\label{sec:discussion}

\subsection{Consistency of RVs and phase curve component amplitudes}
\label{sec: phase curve amplitudes}

\setlength{\extrarowheight}{2pt}
\begin{table}[ht!]
\centering
\caption{Stellar parameters adopted for KIC 8121913, KIC 10068024, and KIC 5479689, taken from the Gaia-Kepler Stellar Properties Catalog \citep{Berger2020}.}
\begin{tabular}{l | c c c}
  & KIC 8121913 & KIC 10068024 & KIC 5479689 \\
\hline
$M_{\star} \ (M_{\odot})$ & $1.456^{+0.069}_{-0.15}$ & $1.22^{+0.175}_{-0.139}$ & $0.966^{+0.03}_{-0.04}$ \\
$R_{\star} \ (R_{\odot})$ & $2.233^{+0.054}_{-0.056}$ & $1.574^{+0.525}_{-0.37}$ & $0.894^{+0.017}_{-0.016}$  \\
$T_{\mathrm{eff}}$ (K) & $5975.3^{+110.2}_{-109.4}$ & $6118.4^{+127.2}_{-121.2}$ & $5459.1^{+90.0}_{-85.2}$  \\
Fe/H & $0.252^{+0.128}_{-0.17}$ & $0.01^{+0.153}_{-0.174}$& $0.161^{+0.116}_{-0.115}$ \\
$\log g$ & $3.896^{+0.032}_{-0.05}$ & $4.092^{+0.183}_{-0.195}$& $4.52^{+0.015}_{-0.024}$ \\
\end{tabular}
\label{tab: stellar parameters}
\end{table}

In the previous section, we treated the phase curve component amplitudes as free parameters within the joint RV/phase curve fit. These amplitudes are not truly free parameters, however, since they stem from physical effects in the star-planet system. Thus, the comparison of the phase curve component amplitudes to one another and to the RV semi-amplitudes can inform whether the joint fits are physically consistent. Here we will describe the relation of the amplitudes to physical parameters and assess their consistency. Within this, we will adopt stellar parameters from the Gaia-Kepler Stellar Properties Catalog \citep{Berger2020}; these are provided in Table~\ref{tab: stellar parameters}.

The atmospheric component of optical phase curves, $A_{\mathrm{refl}}$, includes both light from the primary star reflected off the companion, as well as thermal emission from the companion itself. It is generally difficult to separate reflected light from thermal emission, and it depends on the system as to which dominates. The amplitude of the atmospheric component is given by \citep{Shporer17},
\begin{equation}
\begin{split}
A_{\mathrm{refl}} &\approx \frac{\alpha_{\mathrm{refl}}}{10}\left(\frac{R_p}{a}\right)^2\sin{i} \\ 
&\approx 57\alpha_{\mathrm{refl}}\sin{i}\left(\frac{M_{\star}}{M_{\odot}}\right)^{-2/3}\left(\frac{P}{\mathrm{day}}\right)^{-4/3}\left(\frac{R_p}{R_{\mathrm{J}}}\right)^2\mathrm{ppm}.
\label{eq: Arefl}
\end{split}
\end{equation}
Here $\alpha_{\mathrm{refl}}$ is an order-unity coefficient that depends on the planet's albedo and the efficiency of heat redistribution.

In addition to phase modulations from the planet's atmospheric brightness, there are also gravitationally-induced photometric signals from the host star. The phase component due to Doppler beaming (also known as Doppler boosting) stems from periodic blueshifts and redshifts of the stellar spectrum in response to the orbiting planet. The amplitude of the beaming component is given by 
\begin{equation}
\begin{split}
A_{\mathrm{beam}} &\approx \frac{4 K \alpha_{\mathrm{beam}}}{c} \\
&\approx 2.7\alpha_{\mathrm{beam}}\left(\frac{P}{\mathrm{day}}\right)^{-1/3}\left(\frac{M_{\star}}{M_{\odot}}\right)^{-2/3}\left(\frac{M_p\sin{i}}{M_{\mathrm{J}}}\right)\mathrm{ppm}.
\label{eq: Abeam}
\end{split}
\end{equation}
Here, $K$ is the radial velocity semi-amplitude, $c$ is the speed of light and $\alpha_{\mathrm{beam}}$ is an order-unity coefficient equal to the photon-weighted bandpass-integrated beaming factor. 

The second gravitational signal results from the ellipsoidal distortion of the host star due to tidal potential of the companion. This signal is dominated by a $\cos(4\pi\phi)$ term, although a more complete expansion includes smaller amplitude harmonics such as $\cos(2\pi\phi)$ and $\cos(6\pi\phi)$, which we have ignored in equation \ref{eq: F(phi)} for simplicity. The photometric amplitude of the ellipsoidal distortion is approximately 
\begin{equation}
\begin{split}
A_{\mathrm{ellip}} &\approx \frac{M_p\alpha_{\mathrm{ellip}}\sin^2{i}}{M_{\star}}\left(\frac{R_{\star}}{a}\right)^3 \\ 
&\approx 13\alpha_{\mathrm{ellip}}\sin{i}\left(\frac{R_{\star}}{R_{\odot}}\right)^3\left(\frac{M_{\star}}{M_{\odot}}\right)^{-2}\left(\frac{P}{\mathrm{day}}\right)^{-2}\left(\frac{M_p\sin{i}}{M_{\mathrm{J}}}\right)\mathrm{ppm}.
\label{eq: Aellip}
\end{split}
\end{equation}
The $\alpha_{\mathrm{ellip}}$ coefficient is order-unity, and it relates to the stellar limb darkening and gravity darkening,
\begin{equation}
\alpha_{\mathrm{ellip}} = \frac{3(15+u)(1+g)}{20(3-u)},
\end{equation}
where $u$ is the linear limb darkening coefficient and $g$ is the gravity darkening coefficient \citep{claret11}.

To investigate the consistency of the phase curve components and the RV semi-amplitudes for each system, we will compare their respective constraints on $M_p\sin{i}$. The minimum mass $M_p\sin{i}$ is most robustly constrained from the RV semi-amplitude $K$, but $A_{\mathrm{beam}}$ and $A_{\mathrm{ellip}}$ also provide weak estimates. We first solve equations \ref{eq: Abeam} and \ref{eq: Aellip} for $M_p\sin{i}$. Next, we use the posterior samples from the joint fits in Section \ref{sec: joint radial velocity and light curve modeling} and the stellar parameter estimates (Table~\ref{tab: stellar parameters}) to calculate the posterior distributions of $M_p\sin{i}$. Specifically, for the stellar parameters, we sample from distributions of $M_{\star}$ and $R_{\star}$ consistent with the observed means and error estimates in Table \ref{tab: stellar parameters}. {The $A_{\mathrm{ellip}}$ amplitude includes an extra factor of $\sin{i}$ ($A_{\mathrm{ellip}}\propto M_p\sin^2{i}$; equation \ref{eq: Aellip}), so we average over it by assuming an isotropic distribution of the inclination.} As for the order-unity coefficients $\alpha_{\mathrm{beam}}$ and $\alpha_{\mathrm{ellip}}$, we set them to fixed values based on simple linear relationships with $T_{\mathrm{eff}}$, which were shown to be good fits to these coefficients in the \textit{Kepler} bandpass \citep{millholland17}:
\begin{equation}
\begin{split}
\alpha_{\mathrm{beam}}&\approx-(6\times10^{-4} \mathrm{K}^{-1})T_{\mathrm{eff}}+7.2 \\
\alpha_{\mathrm{ellip}}&\approx-(2.2\times10^{-4} \mathrm{K}^{-1})T_{\mathrm{eff}}+2.6.
\end{split}
\end{equation}
This is an appropriate simplification because the uncertainties in the derived $M_p\sin{i}$ are primarily driven by the spread in posterior samples of $A_{\mathrm{beam}}$ and $A_{\mathrm{ellip}}$. Finally, putting all of this together, we calculate the posterior distributions of $M_p\sin{i}$. The resulting means and $1\sigma$ ranges are given in Table \ref{tab: joint fit}.  

We find that the $M_p\sin{i}$ estimates from RV, $A_{\mathrm{beam}}$, and $A_{\mathrm{ellip}}$ are consistent in almost all cases for our three targets. {The one exception arises within the $M_p\sin{i}$ estimates derived from $A_{\mathrm{ellip}}$, which are systematically lower than those derived from $K$ and $A_{\mathrm{beam}}$ in all three cases, up to $\sim 4\sigma$ lower in the case of KIC 8121913. This discrepancy could arise from the extra factor of $\sin{i}$, which we marginalized over. Additionally, it could be related to uncertainties in $\alpha_{\mathrm{ellip}}$, which we ignored. More broadly, however,} this mass discrepancy is common in photometric phase curve analyses and has been pointed out in previous studies (see \citealt{Shporer17} and references therein). An inconsistency between the ellipsoidal and RV-derived masses was found for KOI-74 by \cite{bloemen11}, an eclipsing binary composed by an A-type and a white dwarf star, with the mass ratio derived by RVs being larger than from ellipsoidal modulations, as in the case of KIC 8121913. Although the reason for this discrepancy is not yet understood, several studies have pointed out different possible causes (e.g., \citealt{Kerkwijk10,pfahl08}).

The above analysis provided a consistency check on both the RVs and the beaming and ellipsoidal components of the phase curve. It is also worthwhile to check the reflection phase curve component amplitude, $A_{\mathrm{refl}}$, although this is complicated by the unknown $\alpha_{\mathrm{refl}}$, $R_p$, and $\sin{i}$. One approach is to use the posterior samples of $A_{\mathrm{refl}}$ from Section \ref{sec: joint radial velocity and light curve modeling} to calculate the combination of parameters $\alpha_{\mathrm{refl}}{R_p}^2\sin{i}$ via equation \ref{eq: Arefl}. We can then examine whether the constraints are reasonably consistent with the RVs. The derived values of $\alpha_{\mathrm{refl}}{R_p}^2\sin{i}$ are given in Table \ref{tab: joint fit}. The values for KIC 8121913 and KIC 10068024 are unity or greater, thus consistent with $R_p\gtrsim R_{\mathrm{Jup}}$ planets, whereas KIC 5479689 suggests a somewhat smaller $R_p$ and/or smaller $\sin{i}$. This constraint is independently consistent with the inferred $M_p\sin{i}$ for KIC 5479689, which was found to be smaller than the other two systems.

\section{Summary and Conclusions}
\label{sec:conclusions}

We performed ground-based follow-up observations of ten non-transiting planet candidates detected through their phase curve modulations via a supervised machine learning approach by \cite{millholland17}. Adequate phase coverage was obtained for six of them. High-spatial resolution images and radial velocity monitoring with the CAFE and HERMES instruments was obtained, yielding planet confirmation for three of the candidates (KIC\,8121913\,b, KIC\,10068024\,b, and KIC\,5479689\,b) and adding credence to the planetary nature of the other three (KIC\,5878307\,b, KIC\,11362225\,b, and KIC\,8026887\,b). {These follow-up observations highlight the potential of phase curve modulations as a detection technique for non-transiting planets.}

We find KIC\,10068024\,b is a massive hot-Jupiter ($M_p=2.0\pm0.4~M_{\rm Jup}$) around a F8-G0 type star potentially in the early stages of the subgiant phase. KIC\,5479689\,b is found to be a sub-Jupiter-mass planet ($M_p=0.5^{+0.4}_{-0.1}~M_{\rm Jup}$) orbiting with a 1.7~days orbital period around a main-sequence solar-like star. Finally, we find KIC\,8121913\,b is a Jupiter-mass planet with a RV-derived mass in the Jupiter regime ($M_p=2.1\pm0.4~M_{\rm Jup}$), in agreement with the beaming amplitude of the phase curve. However, in this case, we find a significant discrepancy with the ellipsoidal amplitude, which points towards a lower mass for this planet of $M_p=0.3^{+0.2}_{-0.1}~M_{\rm Jup}$. This discrepancy has already been reported in other systems with phase curve modulations, but its origin remains unknown. Hence, KIC\,8121913\,b, and additional systems like it can help in the interpretation of this discrepancy and the physics behind it.

Additionally, our RV data suggests that the orbital period of the candidate planet in KIC\,8026887 is actually twice the photometric period derived in \cite{millholland17}. Under such an assumption, our RV analysis provides a significant detection with a planet mass corresponding to a massive gaseous giant ($M_p=4.0^{+1.2}_{-0.7}~M_{\rm Jup}$). However, because we cannot obtain a strong joint fit of both the light curve and RV data with the same orbital period, we do not consider this as a confirmed planet. Finally, we obtained an upper limit to the masses of the other two candidates of $8.5~M_{\rm Jup}$ (KIC\,5878307\,b) and $4.4~M_{\rm Jup}$ (KIC\,11362225\,b).

These hot Jupiters are not known to posses nearby planetary companions and are thus of less interest from the perspective of planetary orbital architectures. However, this phase curve detection technique offers the most promising approach for detecting hot Jupiters in multi-planet systems with mutually-inclined orbits \citep{Millholland16}. {Specifically, if we consider a hypothetical system with a known, transiting super-Earth/sub-Neptune on a $\gtrsim 10$ day orbit, we can inspect the stellar light curve for photometric variations from a \textit{non-transiting} hot Jupiter on an interior $\lesssim 5$ day orbit.} As shown by \cite{Millholland16}, the detection of astrometrically induced transit-timing variations for the {exterior} transiting planet may serve as an additional predictor for the existence of the potential non-transiting hot Jupiter. {Systems containing a mutually-inclined hot Jupiter and exterior, small-mass planet} are predicted to exist if some fraction of hot Jupiters form \textit{in situ} \citep{Batygin16}. The search for phase curve detections of non-transiting hot Jupiters {(and the subsequent radial velocity monitoring)} offers a means towards a direct test of this hypothesis.

The targets presented in this paper are the first non-transiting planets to be confirmed after initial detection of their full-phase photometric variations\footnote{{We note that \cite{faigler13} first found Kepler-76 through phase curve modulations, but it was subsequently found to transit its host star.}}. The photometric detection of non-transiting planets will continue to be relevant to exoplanet exploration as new space-based missions are launched. The \textit{Kepler} mission provided several candidates in this regime and many more will come from TESS \citep{ricker14} and the future PLATO mission \citep{rauer14}. Currently, the \textit{Cheops} mission through one of its working packages is also focused on the study of phase curves through high-precision and high-cadence observations. The targets presented here, and especially KIC\,8121913\,b, offer excellent opportunities to probe hot Jupiter atmospheres and understand the possible discrepancies between RV-derived and photometric masses. Moreover, the method as a whole can be used as a test of theories of hot Jupiter formation, which is still one of the biggest open questions in the field of exoplanets today.


\begin{acknowledgements}
J.L-B acknowledges financial support received from ”la Caixa” Foundation (ID 100010434) and from the European Union’s Horizon 2020 research and innovation programme under the Marie Skłodowska-Curie grant agreement No 847648, with fellowship code LCF/BQ/PI20/11760023. This research has also been partly funded by the Spanish State Research Agency (AEI) Projects No.ESP2017-87676-C5-1-R and No. MDM-2017-0737 Unidad de Excelencia "Mar\'ia de Maeztu"- Centro de Astrobiolog\'ia (INTA-CSIC). S.M. was supported by NASA through the NASA Hubble Fellowship grant \#HST-HF2-51465 awarded by the Space Telescope Science Institute, which is operated by the Association of Universities for Research in Astronomy, Inc., for NASA, under contract NAS5-26555. S.M. was also supported by the NSF Graduate Research Fellowship Program under Grant DGE-1122492. Partly based on observations obtained with the HERMES spectrograph, which is supported by the Research Foundation - Flanders (FWO), Belgium, the Research Council of KU Leuven, Belgium, the Fonds National de la Recherche Scientifique (F.R.S.-FNRS), Belgium, the Royal Observatory of Belgium, the Observatoire de Genève, Switzerland and the Thüringer Landessternwarte Tautenburg, Germany.
\end{acknowledgements}

\bibliographystyle{aa} 
\bibliography{biblio2.bib} 

\begin{thebibliography}{49}
\expandafter\ifx\csname natexlab\endcsname\relax\def\natexlab#1{#1}\fi

\bibitem[{{Aceituno} {et~al.}(2013){Aceituno}, {S{\'a}nchez}, {Grupp}, {Lillo},
  {Hern{\'a}n-Obispo}, {Benitez}, {Montoya}, {Thiele}, {Pedraz}, {Barrado},
  {Dreizler}, \& {Bean}}]{aceituno13}
{Aceituno}, J., {S{\'a}nchez}, S.~F., {Grupp}, F., {et~al.} 2013, \aap, 552,
  A31

\bibitem[{{Akeson} {et~al.}(2013){Akeson}, {Chen}, {Ciardi}, {Crane}, {Good},
  {Harbut}, {Jackson}, {Kane}, {Laity}, {Leifer}, {Lynn}, {McElroy}, {Papin},
  {Plavchan}, {Ram{\'{\i}}rez}, {Rey}, {von Braun}, {Wittman}, {Abajian},
  {Ali}, {Beichman}, {Beekley}, {Berriman}, {Berukoff}, {Bryden}, {Chan},
  {Groom}, {Lau}, {Payne}, {Regelson}, {Saucedo}, {Schmitz}, {Stauffer},
  {Wyatt}, \& {Zhang}}]{akeson13}
{Akeson}, R.~L., {Chen}, X., {Ciardi}, D., {et~al.} 2013, \pasp, 125, 989

\bibitem[{{Aller} {et~al.}(2018){Aller}, {Lillo-Box}, {Vu{\v c}kovi{\'c}}, {Van
  Winckel}, {Jones}, {Montesinos}, {Zorotovic}, \& {Miranda}}]{aller18}
{Aller}, A., {Lillo-Box}, J., {Vu{\v c}kovi{\'c}}, M., {et~al.} 2018, \mnras,
  476, 1140

\bibitem[{{Armstrong} {et~al.}(2020){Armstrong}, {Lopez}, {Adibekyan}, {Booth},
  {Bryant}, {Collins}, {Emsenhuber}, {Huang}, {King}, {Lillo-box}, {Lissauer},
  {Matthews}, {Mousis}, {Nielsen}, {Osborn}, {Otegi}, {Santos}, {Sousa},
  {Stassun}, {Veras}, {Ziegler}, {Acton}, {Almenara}, {Anderson}, {Barrado},
  {Barros}, {Bayliss}, {Belardi}, {Bouchy}, {Briceno}, {Brogi}, {Brown},
  {Burleigh}, {Casewell}, {Chaushev}, {Ciardi}, {Collins}, {Col{\'o}n},
  {Cooke}, {Crossfield}, {D{\'\i}az}, {Deleuil}, {Delgado Mena}, {Demangeon},
  {Dorn}, {Dumusque}, {Eigmuller}, {Fausnaugh}, {Figueira}, {Gan}, {Gand hi},
  {Gill}, {Goad}, {Guenther}, {Helled}, {Hojjatpanah}, {Howell}, {Jackman},
  {Jenkins}, {Jenkins}, {Jensen}, {Kennedy}, {Latham}, {Law}, {Lendl},
  {Lozovsky}, {Mann}, {Moyano}, {McCormac}, {Meru}, {Mordasini}, {Osborn},
  {Pollacco}, {Queloz}, {Raynard}, {Ricker}, {Rowden}, {Santerne}, {Schlieder},
  {Seager}, {Sha}, {Tan}, {Tilbrook}, {Ting}, {Udry}, {Vanderspek}, {Watson},
  {West}, {Wilson}, {Winn}, {Wheatley}, {Villasenor}, {Vines}, \&
  {Zhan}}]{armstrong20}
{Armstrong}, D.~J., {Lopez}, T.~A., {Adibekyan}, V., {et~al.} 2020, arXiv
  e-prints, arXiv:2003.10314

\bibitem[{{Auvergne} {et~al.}(2009){Auvergne}, {Bodin}, {Boisnard}, {Buey},
  {Chaintreuil}, {Epstein}, {Jouret}, {Lam-Trong}, {Levacher}, {Magnan},
  {Perez}, {Plasson}, {Plesseria}, {Peter}, {Steller}, {Tiph{\`e}ne}, {Baglin},
  {Agogu{\'e}}, {Appourchaux}, {Barbet}, {Beaufort}, {Bellenger}, {Berlin},
  {Bernardi}, {Blouin}, {Boumier}, {Bonneau}, {Briet}, {Butler}, {Cautain},
  {Chiavassa}, {Costes}, {Cuvilho}, {Cunha-Parro}, {de Oliveira Fialho},
  {Decaudin}, {Defise}, {Djalal}, {Docclo}, {Drummond}, {Dupuis}, {Exil},
  {Faur{\'e}}, {Gaboriaud}, {Gamet}, {Gavalda}, {Grolleau}, {Gueguen},
  {Guivarc'h}, {Guterman}, {Hasiba}, {Huntzinger}, {Hustaix}, {Imbert},
  {Jeanville}, {Johlander}, {Jorda}, {Journoud}, {Karioty}, {Kerjean},
  {Lafond}, {Lapeyrere}, {Landiech}, {Larqu{\'e}}, {Laudet}, {Le Merrer},
  {Leporati}, {Leruyet}, {Levieuge}, {Llebaria}, {Martin}, {Mazy}, {Mesnager},
  {Michel}, {Moalic}, {Monjoin}, {Naudet}, {Neukirchner}, {Nguyen-Kim},
  {Ollivier}, {Orcesi}, {Ottacher}, {Oulali}, {Parisot}, {Perruchot},
  {Piacentino}, {Pinheiro da Silva}, {Platzer}, {Pontet}, {Pradines},
  {Quentin}, {Rohbeck}, {Rolland}, {Rollenhagen}, {Romagnan}, {Russ}, {Samadi},
  {Schmidt}, {Schwartz}, {Sebbag}, {Smit}, {Sunter}, {Tello}, {Toulouse},
  {Ulmer}, {Vandermarcq}, {Vergnault}, {Wallner}, {Waultier}, \&
  {Zanatta}}]{auvergne09}
{Auvergne}, M., {Bodin}, P., {Boisnard}, L., {et~al.} 2009, \aap, 506, 411

\bibitem[{{Baranne} {et~al.}(1996){Baranne}, {Queloz}, {Mayor}, {Adrianzyk},
  {Knispel}, {Kohler}, {Lacroix}, {Meunier}, {Rimbaud}, \& {Vin}}]{baranne96}
{Baranne}, A., {Queloz}, D., {Mayor}, M., {et~al.} 1996, \aaps, 119, 373

\bibitem[{{Barros} {et~al.}(2017){Barros}, {Gosselin}, {Lillo-Box}, {Bayliss},
  {Delgado Mena}, {Brugger}, {Santerne}, {Armstrong}, {Adibekyan}, {Armstrong},
  {Barrado}, {Bento}, {Boisse}, {Bonomo}, {Bouchy}, {Brown}, {Cochran},
  {Collier Cameron}, {Deleuil}, {Demangeon}, {D{\'\i}az}, {Doyle}, {Dumusque},
  {Ehrenreich}, {Espinoza}, {Faedi}, {Faria}, {Figueira}, {Foxell},
  {H{\'e}brard}, {Hojjatpanah}, {Jackman}, {Lendl}, {Ligi}, {Lovis}, {Melo},
  {Mousis}, {Neal}, {Osborn}, {Pollacco}, {Santos}, {Sefako}, {Shporer},
  {Sousa}, {Triaud}, {Udry}, {Vigan}, \& {Wyttenbach}}]{barros17}
{Barros}, S.~C.~C., {Gosselin}, H., {Lillo-Box}, J., {et~al.} 2017, \aap, 608,
  A25

\bibitem[{{Batygin} {et~al.}(2016){Batygin}, {Bodenheimer}, \&
  {Laughlin}}]{Batygin16}
{Batygin}, K., {Bodenheimer}, P.~H., \& {Laughlin}, G.~P. 2016, \apj, 829, 114

\bibitem[{{Berger} {et~al.}(2020){Berger}, {Huber}, {van Saders}, {Gaidos},
  {Tayar}, \& {Kraus}}]{Berger2020}
{Berger}, T.~A., {Huber}, D., {van Saders}, J.~L., {et~al.} 2020, \aj, 159, 280

\bibitem[{{Bloemen} {et~al.}(2011){Bloemen}, {Marsh}, {{\O}stensen},
  {Charpinet}, {Fontaine}, {Degroote}, {Heber}, {Kawaler}, {Aerts}, {Green},
  {Telting}, {Brassard}, {G{\"a}nsicke}, {Handler}, {Kurtz}, {Silvotti}, {Van
  Grootel}, {Lindberg}, {Pursimo}, {Wilson}, {Gilliland}, {Kjeldsen},
  {Christensen-Dalsgaard}, {Borucki}, {Koch}, {Jenkins}, \&
  {Klaus}}]{bloemen11}
{Bloemen}, S., {Marsh}, T.~R., {{\O}stensen}, R.~H., {et~al.} 2011, \mnras,
  410, 1787

\bibitem[{{Bluhm} {et~al.}(2020){Bluhm}, {Luque}, {Espinoza}, {Pall{\'e}},
  {Caballero}, {Dreizler}, {Livingston}, {Mathur}, {Quirrenbach}, {Stock}, {Van
  Eylen}, {Nowak}, {L{\'o}pez}, {Csizmadia}, {Zapatero Osorio}, {Sch{\"o}fer},
  {Lillo-Box}, {Oshagh}, {Gonz{\'a}lez-{\'A}lvarez}, {Amado}, {Barrado},
  {B{\'e}jar}, {Cale}, {Chaturvedi}, {Cifuentes}, {Cochran}, {Collins},
  {Collins}, {Cort{\'e}s-Contreras}, {D{\'\i}ez Alonso}, {El Mufti},
  {Ercolino}, {Fridlund}, {Gaidos}, {Garc{\'\i}a}, {Georgieva},
  {Gonz{\'a}lez-Cuesta}, {Guerra}, {Hatzes}, {Henning}, {Herrero}, {Hidalgo},
  {Isopi}, {Jeffers}, {Jenkins}, {Jensen}, {K{\'a}bath}, {Kaminski}, {Kemmer},
  {Korth}, {Kossakowski}, {K{\"u}rster}, {Lafarga}, {Mallia}, {Montes},
  {Morales}, {Morales-Calder{\'o}n}, {Murgas}, {Narita}, {Passegger}, {Pedraz},
  {Persson}, {Plavchan}, {Rauer}, {Redfield}, {Reffert}, {Reiners}, {Ribas},
  {Ricker}, {Rodr{\'\i}guez-L{\'o}pez}, {Santos}, {Seager}, {Schlecker},
  {Schweitzer}, {Shan}, {Soto}, {Subjak}, {Tal-Or}, {Trifonov}, {Vanaverbeke},
  {Vanderspek}, {Wittrock}, {Zechmeister}, \& {Zohrabi}}]{bluhm20}
{Bluhm}, P., {Luque}, R., {Espinoza}, N., {et~al.} 2020, \aap, 639, A132

\bibitem[{{Borucki} {et~al.}(2010){Borucki}, {Koch}, {Basri}, {Batalha},
  {Brown}, {Caldwell}, {Caldwell}, {Christensen-Dalsgaard}, {Cochran},
  {DeVore}, {Dunham}, {Dupree}, {Gautier}, {Geary}, {Gilliland}, {Gould},
  {Howell}, {Jenkins}, {Kondo}, {Latham}, {Marcy}, {Meibom}, {Kjeldsen},
  {Lissauer}, {Monet}, {Morrison}, {Sasselov}, {Tarter}, {Boss}, {Brownlee},
  {Owen}, {Buzasi}, {Charbonneau}, {Doyle}, {Fortney}, {Ford}, {Holman},
  {Seager}, {Steffen}, {Welsh}, {Rowe}, {Anderson}, {Buchhave}, {Ciardi},
  {Walkowicz}, {Sherry}, {Horch}, {Isaacson}, {Everett}, {Fischer}, {Torres},
  {Johnson}, {Endl}, {MacQueen}, {Bryson}, {Dotson}, {Haas}, {Kolodziejczak},
  {Van Cleve}, {Chandrasekaran}, {Twicken}, {Quintana}, {Clarke}, {Allen},
  {Li}, {Wu}, {Tenenbaum}, {Verner}, {Bruhweiler}, {Barnes}, \&
  {Prsa}}]{borucki10}
{Borucki}, W.~J., {Koch}, D., {Basri}, G., {et~al.} 2010, Science, 327, 977

\bibitem[{{Borucki} {et~al.}(2012){Borucki}, {Koch}, {Batalha}, {Bryson},
  {Rowe}, {Fressin}, {Torres}, {Caldwell}, {Christensen-Dalsgaard}, {Cochran},
  {DeVore}, {Gautier}, {Geary}, {Gilliland}, {Gould}, {Howell}, {Jenkins},
  {Latham}, {Lissauer}, {Marcy}, {Sasselov}, {Boss}, {Charbonneau}, {Ciardi},
  {Kaltenegger}, {Doyle}, {Dupree}, {Ford}, {Fortney}, {Holman}, {Steffen},
  {Mullally}, {Still}, {Tarter}, {Ballard}, {Buchhave}, {Carter},
  {Christiansen}, {Demory}, {D{\'e}sert}, {Dressing}, {Endl}, {Fabrycky},
  {Fischer}, {Haas}, {Henze}, {Horch}, {Howard}, {Isaacson}, {Kjeldsen},
  {Johnson}, {Klaus}, {Kolodziejczak}, {Barclay}, {Li}, {Meibom}, {Prsa},
  {Quinn}, {Quintana}, {Robertson}, {Sherry}, {Shporer}, {Tenenbaum},
  {Thompson}, {Twicken}, {Van Cleve}, {Welsh}, {Basu}, {Chaplin}, {Miglio},
  {Kawaler}, {Arentoft}, {Stello}, {Metcalfe}, {Verner}, {Karoff}, {Lundkvist},
  {Lund}, {Handberg}, {Elsworth}, {Hekker}, {Huber}, {Bedding}, \&
  {Rapin}}]{borucki12}
{Borucki}, W.~J., {Koch}, D.~G., {Batalha}, N., {et~al.} 2012, \apj, 745, 120

\bibitem[{{Ciceri} {et~al.}(2015){Ciceri}, {Lillo-Box}, {Southworth},
  {Mancini}, {Henning}, \& {Barrado}}]{ciceri14}
{Ciceri}, S., {Lillo-Box}, J., {Southworth}, J., {et~al.} 2015, \aap, 573, L5

\bibitem[{{Claret} \& {Bloemen}(2011)}]{claret11}
{Claret}, A. \& {Bloemen}, S. 2011, \aap, 529, A75

\bibitem[{{Demory} \& {Seager}(2011)}]{demory11}
{Demory}, B.-O. \& {Seager}, S. 2011, \apjs, 197, 12

\bibitem[{{Doyle} {et~al.}(2011){Doyle}, {Carter}, {Fabrycky}, {Slawson},
  {Howell}, {Winn}, {Orosz}, {Welsh}, {Quinn}, {Latham}, {Torres}, {Buchhave},
  {Marcy}, {Fortney}, {Shporer}, {Ford}, {Lissauer}, {Ragozzine}, {Rucker},
  {Batalha}, {Jenkins}, {Borucki}, {Koch}, {Middour}, {Hall}, {McCauliff},
  {Fanelli}, {Quintana}, {Holman}, {Caldwell}, {Still}, {Stefanik}, {Brown},
  {Esquerdo}, {Tang}, {Furesz}, {Geary}, {Berlind}, {Calkins}, {Short},
  {Steffen}, {Sasselov}, {Dunham}, {Cochran}, {Boss}, {Haas}, {Buzasi}, \&
  {Fischer}}]{doyle11}
{Doyle}, L.~R., {Carter}, J.~A., {Fabrycky}, D.~C., {et~al.} 2011, Science,
  333, 1602

\bibitem[{{Esteves} {et~al.}(2015){Esteves}, {De Mooij}, \&
  {Jayawardhana}}]{esteves14}
{Esteves}, L.~J., {De Mooij}, E.~J.~W., \& {Jayawardhana}, R. 2015, \apj, 804,
  150

\bibitem[{{Faigler} \& {Mazeh}(2011)}]{faigler11}
{Faigler}, S. \& {Mazeh}, T. 2011, \mnras, 415, 3921

\bibitem[{{Faigler} {et~al.}(2013){Faigler}, {Tal-Or}, {Mazeh}, {Latham}, \&
  {Buchhave}}]{faigler13}
{Faigler}, S., {Tal-Or}, L., {Mazeh}, T., {Latham}, D.~W., \& {Buchhave}, L.~A.
  2013, \apj, 771, 26

\bibitem[{{Foreman-Mackey} {et~al.}(2013){Foreman-Mackey}, {Hogg}, {Lang}, \&
  {Goodman}}]{2013PASP..125..306F}
{Foreman-Mackey}, D., {Hogg}, D.~W., {Lang}, D., \& {Goodman}, J. 2013, \pasp,
  125, 306

\bibitem[{{Furlan} {et~al.}(2017){Furlan}, {Ciardi}, {Everett}, {Saylors},
  {Teske}, {Horch}, {Howell}, {van Belle}, {Hirsch}, {Gautier}, {Adams},
  {Barrado}, {Cartier}, {Dressing}, {Dupree}, {Gilliland}, {Lillo-Box},
  {Lucas}, \& {Wang}}]{furlan17}
{Furlan}, E., {Ciardi}, D.~R., {Everett}, M.~E., {et~al.} 2017, \aj, 153, 71

\bibitem[{{Goodman} \& {Weare}(2010)}]{2010CAMCS...5...65G}
{Goodman}, J. \& {Weare}, J. 2010, Communications in Applied Mathematics and
  Computational Science, 5, 65

\bibitem[{{Gray}(2005)}]{gray05}
{Gray}, D.~F. 2005, {The Observation and Analysis of Stellar Photospheres}

\bibitem[{{Hills} \& {Dale}(1974)}]{Hills74}
{Hills}, J.~G. \& {Dale}, T.~M. 1974, \aap, 30, 135

\bibitem[{{Hormuth} {et~al.}(2008){Hormuth}, {Brandner}, {Hippler}, \&
  {Henning}}]{hormuth08}
{Hormuth}, F., {Brandner}, W., {Hippler}, S., \& {Henning}, T. 2008, Journal of
  Physics Conference Series, 131, 012051

\bibitem[{{Howell} {et~al.}(2014){Howell}, {Sobeck}, {Haas}, {Still},
  {Barclay}, {Mullally}, {Troeltzsch}, {Aigrain}, {Bryson}, {Caldwell},
  {Chaplin}, {Cochran}, {Huber}, {Marcy}, {Miglio}, {Najita}, {Smith},
  {Twicken}, \& {Fortney}}]{howell14}
{Howell}, S.~B., {Sobeck}, C., {Haas}, M., {et~al.} 2014, \pasp, 126, 398

\bibitem[{{Lillo-Box} {et~al.}(2020){Lillo-Box}, {Aceituno}, {Pedraz},
  {Bergond}, {Galad{\'\i}-Enr{\'\i}quez}, {Azzaro}, {Arroyo-Torres},
  {Fern{\'a}ndez-Mart{\'\i}n}, {Guijarro}, {Hedrosa}, {Hermelo}, {Hoyo}, \&
  {Mart{\'\i}n-Fern{\'a}ndez}}]{lillo-box20}
{Lillo-Box}, J., {Aceituno}, J., {Pedraz}, S., {et~al.} 2020, \mnras, 491, 4496

\bibitem[{{Lillo-Box} {et~al.}(2014){Lillo-Box}, {Barrado}, {~Moya},
  {Montesinos}, {Montalb{\'a}n}, {Bayo}, {Barbieri}, {R{\'e}gulo}, {Mancini},
  {Bouy}, \& {Henning}}]{lillo-box14}
{Lillo-Box}, J., {Barrado}, D., {~Moya}, A., {et~al.} 2014, \aap, 562, A109

\bibitem[{{Lillo-Box} {et~al.}(2012){Lillo-Box}, {Barrado}, \&
  {Bouy}}]{lillo-box12}
{Lillo-Box}, J., {Barrado}, D., \& {Bouy}, H. 2012, \aap, 546, A10

\bibitem[{{Mathur} {et~al.}(2017){Mathur}, {Huber}, {Batalha}, {Ciardi},
  {Bastien}, {Bieryla}, {Buchhave}, {Cochran}, {Endl}, {Esquerdo}, {Furlan},
  {Howard}, {Howell}, {Isaacson}, {Latham}, {MacQueen}, \& {Silva}}]{Mathur17}
{Mathur}, S., {Huber}, D., {Batalha}, N.~M., {et~al.} 2017, \apjs, 229, 30

\bibitem[{{Mazeh} {et~al.}(2012){Mazeh}, {Nachmani}, {Sokol}, {Faigler}, \&
  {Zucker}}]{mazeh12}
{Mazeh}, T., {Nachmani}, G., {Sokol}, G., {Faigler}, S., \& {Zucker}, S. 2012,
  \aap, 541, A56

\bibitem[{{Millholland} \& {Laughlin}(2017)}]{millholland17}
{Millholland}, S. \& {Laughlin}, G. 2017, \aj, 154, 83

\bibitem[{{Millholland} {et~al.}(2016){Millholland}, {Wang}, \&
  {Laughlin}}]{Millholland16}
{Millholland}, S., {Wang}, S., \& {Laughlin}, G. 2016, \apjl, 823, L7

\bibitem[{{Morris}(1985)}]{morris85}
{Morris}, S.~L. 1985, \apj, 295, 143

\bibitem[{{Pfahl} {et~al.}(2008){Pfahl}, {Arras}, \& {Paxton}}]{pfahl08}
{Pfahl}, E., {Arras}, P., \& {Paxton}, B. 2008, \apj, 679, 783

\bibitem[{{Raskin} {et~al.}(2011){Raskin}, {van Winckel}, {Hensberge},
  {Jorissen}, {Lehmann}, {Waelkens}, {Avila}, {de Cuyper}, {Degroote},
  {Dubosson}, {Dumortier}, {Fr{\'e}mat}, {Laux}, {Michaud}, {Morren}, {Perez
  Padilla}, {Pessemier}, {Prins}, {Smolders}, {van Eck}, \&
  {Winkler}}]{raskin11}
{Raskin}, G., {van Winckel}, H., {Hensberge}, H., {et~al.} 2011, \aap, 526, A69

\bibitem[{{Rauer} {et~al.}(2014){Rauer}, {Catala}, {Aerts}, {Appourchaux},
  {Benz}, {Brandeker}, {Christensen-Dalsgaard}, {Deleuil}, {Gizon}, {Goupil},
  {G{\"u}del}, {Janot-Pacheco}, {Mas-Hesse}, {Pagano}, {Piotto}, {Pollacco},
  {Santos}, {Smith}, {Su{\'a}rez}, {Szab{\'o}}, {Udry}, {Adibekyan}, {Alibert},
  {Almenara}, {Amaro-Seoane}, {Eiff}, {Asplund}, {Antonello}, {Barnes},
  {Baudin}, {Belkacem}, {Bergemann}, {Bihain}, {Birch}, {Bonfils}, {Boisse},
  {Bonomo}, {Borsa}, {Brand{\~a}o}, {Brocato}, {Brun}, {Burleigh}, {Burston},
  {Cabrera}, {Cassisi}, {Chaplin}, {Charpinet}, {Chiappini}, {Church},
  {Csizmadia}, {Cunha}, {Damasso}, {Davies}, {Deeg}, {D{\'{\i}}az}, {Dreizler},
  {Dreyer}, {Eggenberger}, {Ehrenreich}, {Eigm{\"u}ller}, {Erikson}, {Farmer},
  {Feltzing}, {de Oliveira Fialho}, {Figueira}, {Forveille}, {Fridlund},
  {Garc{\'{\i}}a}, {Giommi}, {Giuffrida}, {Godolt}, {Gomes da Silva},
  {Granzer}, {Grenfell}, {Grotsch-Noels}, {G{\"u}nther}, {Haswell}, {Hatzes},
  {H{\'e}brard}, {Hekker}, {Helled}, {Heng}, {Jenkins}, {Johansen},
  {Khodachenko}, {Kislyakova}, {Kley}, {Kolb}, {Krivova}, {Kupka}, {Lammer},
  {Lanza}, {Lebreton}, {Magrin}, {Marcos-Arenal}, {Marrese}, {Marques},
  {Martins}, {Mathis}, {Mathur}, {Messina}, {Miglio}, {Montalban}, {Montalto},
  {Monteiro}, {Moradi}, {Moravveji}, {Mordasini}, {Morel}, {Mortier},
  {Nascimbeni}, {Nelson}, {Nielsen}, {Noack}, {Norton}, {Ofir}, {Oshagh},
  {Ouazzani}, {P{\'a}pics}, {Parro}, {Petit}, {Plez}, {Poretti}, {Quirrenbach},
  {Ragazzoni}, {Raimondo}, {Rainer}, {Reese}, {Redmer}, {Reffert},
  {Rojas-Ayala}, {Roxburgh}, {Salmon}, {Santerne}, {Schneider}, {Schou},
  {Schuh}, {Schunker}, {Silva-Valio}, {Silvotti}, {Skillen}, {Snellen}, {Sohl},
  {Sousa}, {Sozzetti}, {Stello}, {Strassmeier}, {{\v S}vanda}, {Szab{\'o}},
  {Tkachenko}, {Valencia}, {Van Grootel}, {Vauclair}, {Ventura}, {Wagner},
  {Walton}, {Weingrill}, {Werner}, {Wheatley}, \& {Zwintz}}]{rauer14}
{Rauer}, H., {Catala}, C., {Aerts}, C., {et~al.} 2014, Experimental Astronomy,
  38, 249

\bibitem[{{Ricker} {et~al.}(2014){Ricker}, {Winn}, {Vanderspek}, {Latham},
  {Bakos}, {Bean}, {Berta-Thompson}, {Brown}, {Buchhave}, {Butler}, {Butler},
  {Chaplin}, {Charbonneau}, {Christensen-Dalsgaard}, {Clampin}, {Deming},
  {Doty}, {De Lee}, {Dressing}, {Dunham}, {Endl}, {Fressin}, {Ge}, {Henning},
  {Holman}, {Howard}, {Ida}, {Jenkins}, {Jernigan}, {Johnson}, {Kaltenegger},
  {Kawai}, {Kjeldsen}, {Laughlin}, {Levine}, {Lin}, {Lissauer}, {MacQueen},
  {Marcy}, {McCullough}, {Morton}, {Narita}, {Paegert}, {Palle}, {Pepe},
  {Pepper}, {Quirrenbach}, {Rinehart}, {Sasselov}, {Sato}, {Seager},
  {Sozzetti}, {Stassun}, {Sullivan}, {Szentgyorgyi}, {Torres}, {Udry}, \&
  {Villasenor}}]{ricker14}
{Ricker}, G.~R., {Winn}, J.~N., {Vanderspek}, R., {et~al.} 2014, in Society of
  Photo-Optical Instrumentation Engineers (SPIE) Conference Series, Vol. 9143,
  Society of Photo-Optical Instrumentation Engineers (SPIE) Conference Series,
  20

\bibitem[{{Rybicki} \& {Lightman}(1979)}]{Rybicki79}
{Rybicki}, G.~B. \& {Lightman}, A.~P. 1979, {Radiative processes in
  astrophysics}

\bibitem[{{Santerne} {et~al.}(2018){Santerne}, {Brugger}, {Armstrong},
  {Adibekyan}, {Lillo-Box}, {Gosselin}, {Aguichine}, {Almenara}, {Barrado},
  {Barros}, {Bayliss}, {Boisse}, {Bonomo}, {Bouchy}, {Brown}, {Deleuil},
  {Delgado Mena}, {Demangeon}, {D{\'\i}az}, {Doyle}, {Dumusque}, {Faedi},
  {Faria}, {Figueira}, {Foxell}, {Giles}, {H{\'e}brard}, {Hojjatpanah},
  {Hobson}, {Jackman}, {King}, {Kirk}, {Lam}, {Ligi}, {Lovis}, {Louden},
  {McCormac}, {Mousis}, {Neal}, {Osborn}, {Pepe}, {Pollacco}, {Santos},
  {Sousa}, {Udry}, \& {Vigan}}]{santerne18}
{Santerne}, A., {Brugger}, B., {Armstrong}, D.~J., {et~al.} 2018, Nature
  Astronomy, 2, 393

\bibitem[{{Santerne} {et~al.}(2012){Santerne}, {Moutou}, {Barros}, {Damiani},
  {D{\'\i}az}, {Almenara}, {Bonomo}, {Bouchy}, {Deleuil}, \&
  {H{\'e}brard}}]{santerne12a}
{Santerne}, A., {Moutou}, C., {Barros}, S.~C.~C., {et~al.} 2012, \aap, 544, L12

\bibitem[{{Shporer}(2017)}]{Shporer17}
{Shporer}, A. 2017, \pasp, 129, 072001

\bibitem[{{Smith} {et~al.}(2012){Smith}, {Stumpe}, {Van Cleve}, {Jenkins},
  {Barclay}, {Fanelli}, {Girouard}, {Kolodziejczak}, {McCauliff}, {Morris}, \&
  {Twicken}}]{smith12}
{Smith}, J.~C., {Stumpe}, M.~C., {Van Cleve}, J.~E., {et~al.} 2012, \pasp, 124,
  1000

\bibitem[{{Soto} {et~al.}(2021){Soto}, {Anglada-Escud{\'e}}, {Dreizler},
  {Molaverdikhani}, {Kemmer}, {Rodr{\'\i}guez-L{\'o}pez}, {Lillo-Box},
  {Pall{\'e}}, {Espinoza}, {Caballero}, {Quirrenbach}, {Ribas}, {Reiners},
  {Narita}, {Hirano}, {Amado}, {B{\'e}jar}, {Bluhm}, {Burke}, {Caldwell},
  {Charbonneau}, {Cloutier}, {Collins}, {Cort{\'e}s-Contreras}, {Girardin},
  {Guerra}, {Harakawa}, {Hatzes}, {Irwin}, {Jenkins}, {Jensen}, {Kawauchi},
  {Koyati}, {Kudo}, {Kunimoto}, {Kuzuhara}, {Latham}, {Montes}, {Morales},
  {Mori}, {Nelson}, {Omiya}, {Pedraz}, {Passegger}, {Rackham}, {Rudat},
  {Schlieder}, {Sch{\"o}fer}, {Schweitzer}, {Selezneva}, {Stockdale}, {Tamura},
  {Trifonov}, {Vanderspek}, \& {Watanabe}}]{soto21}
{Soto}, M.~G., {Anglada-Escud{\'e}}, G., {Dreizler}, S., {et~al.} 2021, arXiv
  e-prints, arXiv:2102.11640

\bibitem[{{Strehl}(1902)}]{strehl1902}
{Strehl}, K. 1902, Astronomische Nachrichten, 158, 89

\bibitem[{{Stumpe} {et~al.}(2014){Stumpe}, {Smith}, {Catanzarite}, {Van Cleve},
  {Jenkins}, {Twicken}, \& {Girouard}}]{Stumpe14}
{Stumpe}, M.~C., {Smith}, J.~C., {Catanzarite}, J.~H., {et~al.} 2014, \pasp,
  126, 100

\bibitem[{{Stumpe} {et~al.}(2012){Stumpe}, {Smith}, {Van Cleve}, {Twicken},
  {Barclay}, {Fanelli}, {Girouard}, {Jenkins}, {Kolodziejczak}, {McCauliff}, \&
  {Morris}}]{stumpe12}
{Stumpe}, M.~C., {Smith}, J.~C., {Van Cleve}, J.~E., {et~al.} 2012, \pasp, 124,
  985

\bibitem[{{van Kerkwijk} {et~al.}(2010){van Kerkwijk}, {Rappaport}, {Breton},
  {Justham}, {Podsiadlowski}, \& {Han}}]{Kerkwijk10}
{van Kerkwijk}, M.~H., {Rappaport}, S.~A., {Breton}, R.~P., {et~al.} 2010,
  \apj, 715, 51

\end{thebibliography}

\newpage




\appendix
\newpage
\section{Tables}
\begin{table*}
\caption{Cross-correlation funtion properties for KIC5479689.}
\label{tab:RV_start}
\begin{tabular}{ccccc}
\hline \hline
JD & RV (km/s) & eRV (km/s) & FWHM (km/s) & Inst. \\
\hline
2458645.652628276 & 8.117 & 0.064 & 8.0 & CAFE \\
2458646.513237323 & 8.69 & 0.10 & 8.3 & CAFE \\
2458674.652914972 & 7.47 & 0.18 & 7.8 & CAFE \\
2458675.518614139 & 7.763 & 0.075 & 8.0 & CAFE \\
2458675.627853993 & 7.755 & 0.057 & 7.7 & CAFE \\
2458701.433922937 & 7.752 & 0.042 & 7.2 & CAFE \\
2458701.550056568 & 7.618 & 0.071 & 7.7 & CAFE \\
2459046.493358 & 7.747 & 0.018 & 7.1 & HERMES \\
2459046.6134008 & 7.098 & 0.022 & 10.0 & HERMES \\
2459049.6490909 & 6.998 & 0.019 & 6.5 & HERMES \\
2459050.4994282 & 7.430 & 0.017 & 7.2 & HERMES \\
\hline
\end{tabular}
\end{table*}

\begin{table*}
\caption{Cross-correlation funtion properties for KIC5878307.}
\begin{tabular}{ccccc}
\hline \hline
JD & RV (km/s) & eRV (km/s) & FWHM (km/s) & Inst. \\
\hline
2458645.53556366 & -25.42 & 0.26 & 10.3 & CAFE \\
2458674.524131845 & -25.50 & 0.19 & 11.3 & CAFE \\
\hline
\end{tabular}
\end{table*}

\begin{table*}
\caption{Cross-correlation funtion properties for KIC8026887.}
\begin{tabular}{ccccc}
\hline \hline
JD & RV (km/s) & eRV (km/s) & FWHM (km/s) & Inst. \\
\hline
2458645.611684904 & 9.320 & 0.090 & 9.2 & CAFE \\
2458646.451271749 & 9.289 & 0.096 & 6.4 & CAFE \\
2458674.611682581 & 8.54 & 0.11 & 9.2 & CAFE \\
2458675.47803279 & 8.786 & 0.049 & 9.1 & CAFE \\
2458675.585808265 & 8.248 & 0.086 & 9.1 & CAFE \\
2458701.511810438 & 8.07 & 0.12 & 8.5 & CAFE \\
2458701.686715424 & 7.84 & 0.28 & 8.2 & CAFE \\
2459047.5033592 & 8.552 & 0.022 & 7.2 & HERMES \\
2459050.6230106 & 8.983 & 0.020 & 7.8 & HERMES \\
\hline
\end{tabular}
\end{table*}

\begin{table*}
\caption{Cross-correlation funtion properties for KIC8121913.}
\begin{tabular}{ccccc}
\hline \hline
JD & RV (km/s) & eRV (km/s) & FWHM (km/s) & Inst. \\
\hline
2458645.450517782 & -23.483 & 0.075 & 11.5 & CAFE \\
2458646.412329766 & -23.359 & 0.064 & 9.7 & CAFE \\
2458674.442495487 & -22.996 & 0.040 & 11.0 & CAFE \\
2458675.439193288 & -23.460 & 0.046 & 10.6 & CAFE \\
2458675.547551508 & -22.940 & 0.043 & 10.6 & CAFE \\
2458701.396477561 & -23.354 & 0.043 & 10.9 & CAFE \\
2458617.579298275 & -23.098 & 0.044 & 10.8 & CAFE \\
2458617.594995345 & -22.850 & 0.042 & 10.8 & CAFE \\
2458618.558319524 & -23.032 & 0.035 & 11.3 & CAFE \\
2458618.631484145 & -23.159 & 0.037 & 11.2 & CAFE \\
2458626.556106488 & -23.161 & 0.046 & 11.0 & CAFE \\
2458626.57112051 & -23.336 & 0.080 & 10.9 & CAFE \\
2458627.631194962 & -22.944 & 0.094 & 11.3 & CAFE \\
2459024.455894218 & -23.007 & 0.048 & 10.9 & CAFE \\
2459024.583168107 & -23.411 & 0.062 & 10.5 & CAFE \\
2459025.468880206 & -23.193 & 0.043 & 10.7 & CAFE \\
2459025.586753734 & -23.311 & 0.043 & 11.2 & CAFE \\
2459046.4690974 & -23.263 & 0.014 & 11.1 & HERMES \\
2459046.7003785 & -23.327 & 0.016 & 11.9 & HERMES \\
2459047.5290948 & -22.968 & 0.013 & 11.2 & HERMES \\
2459047.6427982 & -24.361 & 0.022 & 12.8 & HERMES \\
2459049.4647511 & -23.142 & 0.015 & 11.7 & HERMES \\
2459049.6083404 & -23.458 & 0.019 & 10.3 & HERMES \\
2459050.4744119 & -23.703 & 0.015 & 11.7 & HERMES \\
2459050.656347 & -23.407 & 0.014 & 13.4 & HERMES \\
\hline
\end{tabular}
\end{table*}

\begin{table*}
\caption{Cross-correlation funtion properties for KIC10068024.}
\begin{tabular}{ccccc}
\hline \hline
JD & RV (km/s) & eRV (km/s) & FWHM (km/s) & Inst. \\
\hline
2458645.416993691 & -22.23 & 0.18 & 12.3 & CAFE \\
2458646.544204343 & -22.527 & 0.096 & 9.5 & CAFE \\
2458674.412376203 & -23.261 & 0.062 & 11.5 & CAFE \\
2458675.408493705 & -22.716 & 0.077 & 11.1 & CAFE \\
2458675.656267811 & -23.55 & 0.15 & 11.2 & CAFE \\
2458701.577313185 & -23.161 & 0.060 & 11.2 & CAFE \\
2458701.642292621 & -23.42 & 0.13 & 10.9 & CAFE \\
2458702.549648101 & -22.795 & 0.067 & 12.0 & CAFE \\
2459046.5195146 & -23.501 & 0.021 & 9.3 & HERMES \\
2459047.5494006 & -23.952 & 0.024 & 13.1 & HERMES \\
2459047.6733413 & -23.343 & 0.034 & 12.5 & HERMES \\
2459049.5179242 & -24.418 & 0.026 & 8.7 & HERMES \\
\hline
\end{tabular}
\end{table*}

\begin{table*}
\caption{Cross-correlation funtion properties for KIC11362225.}
\label{tab:RV_end}
\begin{tabular}{ccccc}
\hline \hline
JD & RV (km/s) & eRV (km/s) & FWHM (km/s) & Inst. \\
\hline
2458617.613913064 & -19.18 & 0.38 & 80.0 & CAFE \\
2458617.628924661 & -19.22 & 0.43 & 85.3 & CAFE \\
2458618.578316415 & -19.10 & 0.20 & 73.8 & CAFE \\
2458618.653932977 & -19.21 & 0.32 & 70.5 & CAFE \\
2458626.596599354 & -19.31 & 0.34 & 78.8 & CAFE \\
2458626.611615575 & -19.32 & 0.25 & 75.4 & CAFE \\
2458627.653016619 & -19.46 & 0.35 & 71.3 & CAFE \\
\hline
\end{tabular}
\end{table*}

\begin{table*}
\caption{Prior and posterior distributions for the analysis of CAFE radial velocities of KIC8121913 from Sect.~\ref{sec:cafe}.}
\label{tab:RVposterior1}
\begin{tabular}{lcc}
\hline
Parameter & Priors & Posteriors \\
\hline
Orbital period, $P_b$ [days] & $\mathcal{G}$(3.2943,0.0029) & $3.29444^{+0.00014}_{-0.00015}$ \\
Time of mid-transit, $T_{\rm 0,b}-2400000$ [days] & $\mathcal{G}$(56427.86897,0.001) & $56427.8689^{+0.0011}_{-0.0010}$ \\
RV semi-amplitude, $K_{\rm b}$ [m/s] & $\mathcal{U}$(0.0,1000.0) & $252^{+53}_{-41}$ \\
$\delta_{\rm CAFE2}$ [km/s] & $\mathcal{U}$(-30.0,-20.0) & $-23.172^{+0.035}_{-0.035}$ \\
$\sigma_{\rm CAFE2}$ [m/s] & $\mathcal{U}$(0.0,0.5) & $103^{+42}_{-27}$ \\
Planet mass, $m_{b}\sin{i_b}$ [$M_{\oplus}$] & (derived) & $780^{+170}_{-160}$ \\
\hline
\end{tabular}
\end{table*}

\begin{table*}
\caption{Prior and posterior distributions for the analysis of CAFE radial velocities of KIC10068024 from Sect.~\ref{sec:cafe}.}
\begin{tabular}{lcc}
\hline
Parameter & Priors & Posteriors \\
\hline
Orbital period, $P_b$ [days] & $\mathcal{G}$(2.0735,0.0012) & $2.07351^{+0.00030}_{-0.00024}$ \\
Time of mid-transit, $T_{\rm 0,b}-2400000$ [days] & $\mathcal{G}$(56424.26451,0.001) & $56424.26452^{+0.00096}_{-0.0011}$ \\
RV semi-amplitude, $K_{\rm b}$ [m/s] & $\mathcal{U}$(0.0,1000.0) & $344^{+96}_{-91}$ \\
$\delta_{\rm CAFE2}$ [km/s] & $\mathcal{U}$(-27.0,-17.0) & $-22.984^{+0.086}_{-0.064}$ \\
$\sigma_{\rm CAFE2}$ [m/s] & $\mathcal{U}$(0.0,0.5) & $151^{+130}_{-87}$ \\
Planet mass, $m_{b}\sin{i_b}$ [$M_{\oplus}$] & (derived) & $750^{+290}_{-200}$ \\
\hline
\end{tabular}
\end{table*}

\begin{table*}
\caption{Prior and posterior distributions for the analysis of CAFE radial velocities of KIC5479689 from Sect.~\ref{sec:cafe}.}
\begin{tabular}{lcc}
\hline
Parameter & Priors & Posteriors \\
\hline
Orbital period, $P_b$ [days] & $\mathcal{G}$(1.7012,0.0008) & $1.70139^{+0.00045}_{-0.00035}$ \\
Time of mid-transit, $T_{\rm 0,b}-2400000$ [days] & $\mathcal{G}$(56423.38036,0.001) & $56423.38026^{+0.0010}_{-0.00099}$ \\
RV semi-amplitude, $K_{\rm b}$ [m/s] & $\mathcal{U}$(0.0,1000.0) & $239^{+48}_{-82}$ \\
$\delta_{\rm CAFE2}$ [m/s] & $\mathcal{U}$(-10.0,10.0) & $-53^{+73}_{-64}$ \\
$\sigma_{\rm CAFE2}$ [m/s] & $\mathcal{U}$(0.0,0.5) & $84^{+93}_{-64}$ \\
Planet mass, $m_{b}\sin{i_b}$ [M$_{\oplus}$] & (derived) & $400^{+150}_{-120}$ \\
\hline
\end{tabular}
\end{table*}

\begin{table*}
\caption{Prior and posterior distributions for the analysis of CAFE radial velocities of KIC8026887 from Sect.~\ref{sec:cafe}.}
\begin{tabular}{lcc}
\hline
Parameter & Priors & Posteriors \\
\hline
Orbital period, $P_b$ [days] & $\mathcal{G}$(3.8464,0.0009) & $3.84558^{+0.00029}_{-0.00029}$ \\
Time of mid-transit, $T_{\rm 0,b}-2400000$ [days] & $\mathcal{G}$(56424.27252,0.001) & $56424.2726^{+0.0010}_{-0.0010}$ \\
RV semi-amplitude, $K_{\rm b}$ [m/s] & $\mathcal{U}$(0.0,1000.0) & $505^{+1.4)e+02}_{-91}$ \\
$\delta_{\rm CAFE2}$ [km/s] & $\mathcal{U}$(0.0,20.0) & $8.78^{+0.11}_{-0.10}$ \\
$\sigma_{\rm CAFE2}$ [m/s] & $\mathcal{U}$(0.0,0.3) & $243^{+39}_{-46}$ \\
$\delta_{\rm HERMES}$ [km/s] & $\mathcal{U}$(0.0,20.0) & $8.89^{+0.17}_{-0.17}$ \\
$\sigma_{\rm HERMES}$ [m/s] & $\mathcal{U}$(0.0,0.3) & $126^{+110}_{-96}$ \\
Planet mass, $m_{b}\sin{i_b}$ [M$_{\oplus}$] & (derived) & $1270^{+330}_{-270}$ \\
\hline
\end{tabular}
\end{table*}

\begin{table*}
\caption{Prior and posterior distributions for the analysis of CAFE radial velocities of KIC11362225 from Sect.~\ref{sec:cafe}.}
\begin{tabular}{lcc}
\hline
Parameter & Priors & Posteriors \\
\hline
Orbital period, $P_b$ [days] & $\mathcal{G}$(2.7513,0.0021) & $2.7512^{+0.0025}_{-0.0017}$ \\
Time of mid-transit, $T_{\rm 0,b}-2400000$ [days] & $\mathcal{G}$(56423.94574,0.001) & $56423.9458^{+0.0010}_{-0.0011}$ \\
RV semi-amplitude, $K_{\rm b}$ [m/s] & $\mathcal{U}$(0.0,1000.0) & $(170^{+140}_{-120}$ \\
$\delta_{\rm CAFE2}$ [km/s] & $\mathcal{U}$(-27.0,-15.0) & $-19.27^{+0.15}_{-0.14}$ \\
$\sigma_{\rm CAFE2}$ [m/s] & $\mathcal{U}$(0.0,0.3) & $99^{+110}_{-68}$ \\
Planet mass, $m_{b}\sin{i_b}$ [M$_{\oplus}$] & (derived) & $310^{+360}_{-220}$ \\
\hline
\end{tabular}
\end{table*}

\begin{table*}
\caption{Prior and posterior distributions for the analysis of CAFE radial velocities of KIC5878307 from Sect.~\ref{sec:cafe}.}
\label{tab:RVposterior2}
\begin{tabular}{lcc}
\hline
Parameter & Priors & Posteriors \\
\hline
Orbital period, $P_b$ [days] & $\mathcal{G}$(2.0466,0.0013) & $2.0465^{+0.0013}_{-0.0012}$ \\
Time of mid-transit, $T_{\rm 0,b}-2400000$ [days] & $\mathcal{G}$(56424.92085,0.1) & $56424.923^{+0.099}_{-0.098}$ \\
RV semi-amplitude, $K_{\rm b}$ [m/s] & $\mathcal{U}$(0.0,1000.0) & $370^{+290}_{-260}$ \\
$\delta_{\rm CAFE2}$ [km/s] & $\mathcal{U}$(-26.0,-24.0) & $-25.42^{+0.37}_{-0.33}$ \\
$\sigma_{\rm CAFE2}$ [m/s] & $\mathcal{U}$(0.0,0.3) & $134^{+110}_{-95}$ \\
Planet mass, $m_{b}\sin{i_b}$ [M$_{\oplus}$] & (derived) & $690^{+660}_{-480}$ \\
\hline
\end{tabular}
\end{table*}





\newpage
\section{Posterior distributions from the joint radial velocity and light curve fit}
\label{sec: posterior distributions}

\begin{figure*}
\centering
\includegraphics[width=\textwidth]{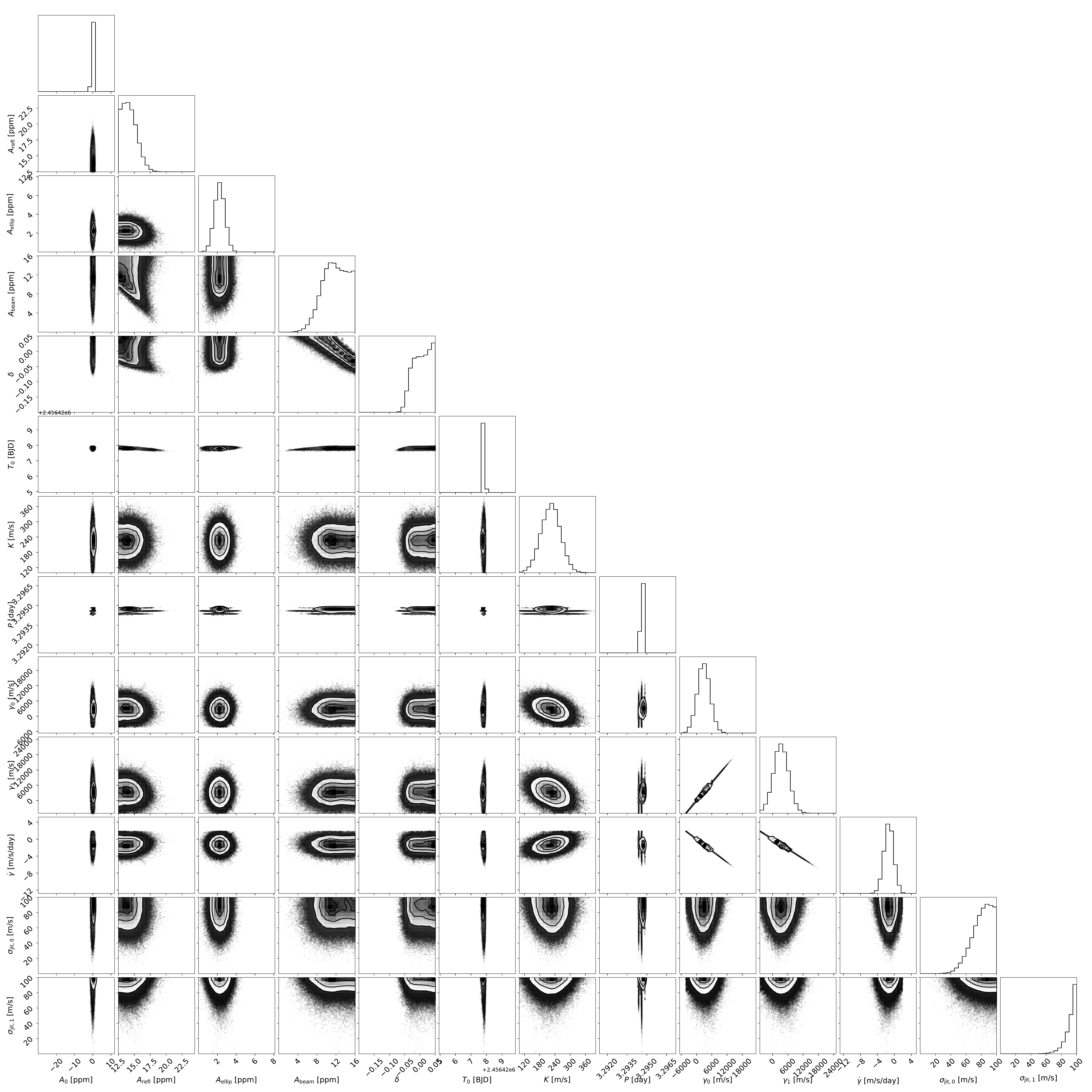}
\caption{Corner plot of the joint posterior parameter distributions for KIC 8121913.}
\label{fig: corner plots}
\end{figure*}

\begin{figure*}
\centering
\includegraphics[width=\textwidth]{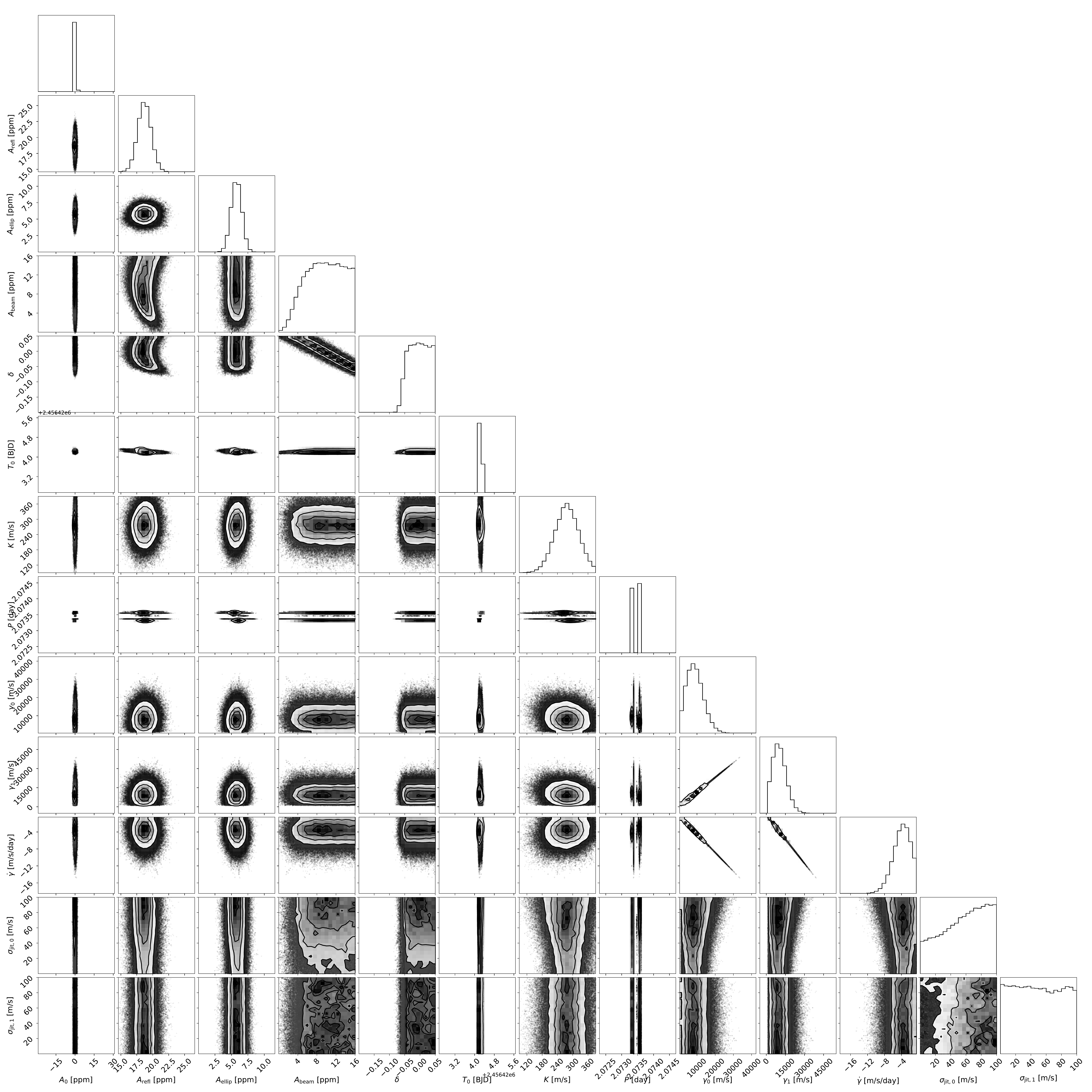}
\caption{Corner plot of the joint posterior parameter distributions for KIC 10068024.}
\label{fig: corner plots}
\end{figure*}

\begin{figure*}
\centering
\includegraphics[width=\textwidth]{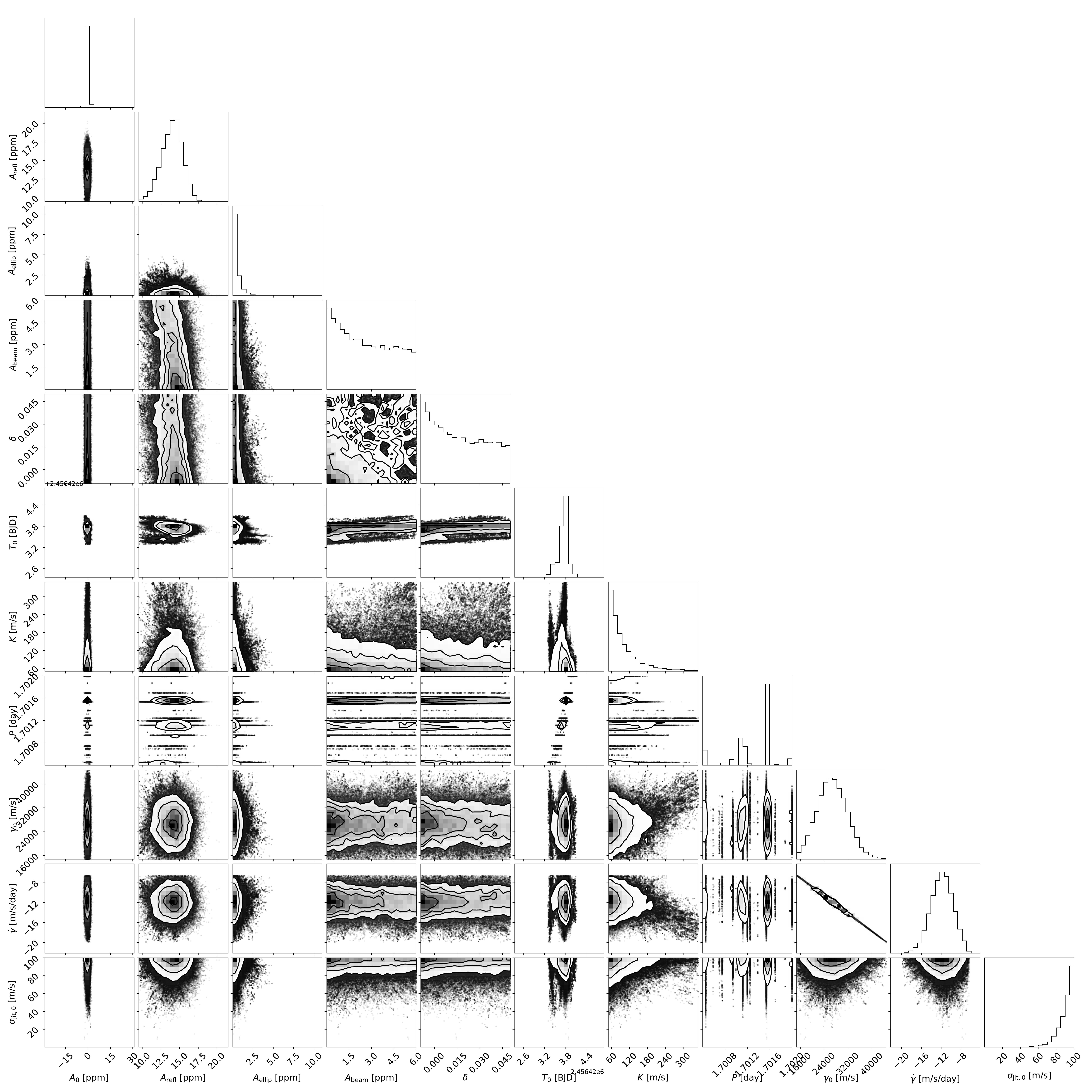}
\caption{Corner plot of the joint posterior parameter distributions for KIC 5479689.}
\label{fig: corner plots}
\end{figure*}

\end{document}